# AI-powered software testing tools: A systematic review and empirical assessment of their features and limitations

| | | |
|---|---|---|
| Vahid Garousi<br>Queen's University Belfast, UK<br>Azerbaijan Technical University, Azerbaijan<br>v.garousi@qub.ac.uk | Nithin Joy<br>British Telecom PLC, UK<br>nithin.joy@bt.com | Zafar Jafarov<br>Department of Information Technologies,<br>Azerbaijan Technical University, Azerbaijan<br>zafar.cafarov@aztu.edu.az |
| Alper Buğra Keleş, Sevde Değirmenci, Ece Özdemir<br>Testinium A.Ş., Türkiye<br>{alper.keles, sevde.degirmenci, ece.ozdemir}@testinium.com | | Ryan Zarringhalami<br>Bahar Software Engineering Consulting, UK<br>ryan.zarringhalami@gmail.com |

**Abstract:**

*Context:* The rise of Artificial Intelligence (AI) in software engineering has led to the development of AI-powered testing tools aimed at supporting human test engineers and improving the efficiency and effectiveness of software testing. However, a systematic evaluation is needed to better understand their capabilities, benefits, and limitations.

*Objective:* This study has two objectives: (1) to conduct a Systematic Tool Review (STR) to identify and categorize AI-powered testing tools based on their AI-driven features; and (2) to perform an empirical study on two representative AI-powered testing tools to assess their effectiveness, efficiency, and limitations in real-world testing scenarios.

*Method:* Our STR identified 56 AI-powered testing tools from industry sources and categorized them based on features such as self-healing tests, visual testing, and AI-powered test generation. Following the STR, two tools—Parasoft Selenic and SmartBear VisualTest—were selected for empirical evaluation using two open-source industrial software systems. Their performance was compared with traditional test automation to evaluate improvements in efficiency and accuracy.

*Results:* The STR provides a comprehensive taxonomy of current AI-powered tools, revealing adoption trends and feature gaps. The empirical study shows that AI-powered testing can enhance test execution speed and reduce test maintenance effort. However, it also highlights critical limitations, including difficulty handling complex UI changes, lack of contextual understanding, and erroneous outputs (due to AI misclassifications). Such errors in test results are especially problematic, as testing itself is meant to identify defects—defects within the testing process undermine its value. Additionally, human oversight remains essential: testers must invest significant effort in reviewing and correcting AI-generated artifacts, which reduces the net efficiency gains promised by AI-powered tools.

*Conclusion:* While AI-powered testing tools hold great promise, further advancements are needed in AI reliability, domain awareness, and reducing the human effort required to oversee AI-generated outputs.

# Table of Contents

# 1 INTRODUCTION

Software testing is a critical yet resource-intensive activity in software engineering, and is essential for ensuring software quality, reliability, and security. Over the years, testing has evolved from manual execution to scripted automation, facilitated by a large number of test tools and frameworks such as Selenium and JUnit. While conventional test automation has improved effectiveness and efficiency of testing, it still suffers from various issues such as fragile test scripts, high maintenance costs of test code, and limited adaptability to dynamic software environments.

The rise of Artificial Intelligence (AI) in software testing has introduced a new generation of AI-powered testing tools [1-3] that leverage machine learning (ML), natural language processing (NLP), and computer vision to automate key testing tasks such as AI-powered test generation, self-healing automation, visual regression testing, and failure analysis. AI-powered testing tools promise to reduce manual intervention of human test engineers, enhance test accuracy, and optimize test execution workflows.

Industry adoption of AI-powered software testing is accelerating. Market reports such as the one by Gartner Research [4] predict that by 2027, 80% of enterprises will have integrated AI-augmented testing tools. Despite this rapid growth, empirical studies evaluating these tools remain limited. While individual AI-powered tools claim to improve efficiency and defect detection, there is a lack of evaluations comparing their capabilities, benefits, and limitations. Existing scientific literature, e.g., [5, 6], largely focuses on AI techniques for test automation but does not assess how AI-powered tools perform in real-world testing scenarios.

This study aims to bridge the above gap by conducting a Systematic Tool Review (STR) of all AI-powered testing tools, followed by an empirical evaluation of two representative tools. The key objectives are:

- To systematically categorize AI-powered testing tools based on their key capabilities, such as self-healing automation, AI-based test generation, and visual testing.
- To empirically evaluate two AI-powered test automation tools, assessing their effectiveness, efficiency, and limitations when applied to real-world software systems.
- To identify key limitations in AI-powered testing and provide insights for future improvements in AI-based test automation.

This study makes the following contributions:

- Comprehensive STR: A structured analysis of 56 AI-powered testing tools, categorizing their features, adoption trends, and capability gaps.
- Empirical evaluation of AI-powered testing tools: A practical assessment of two AI-powered testing tools, applied to two open-source software projects, to analyze their real-world effectiveness and efficiency.
- Identification of limitations and future research directions: The study highlights challenges in AI-based testing, including false positives, contextual limitations, and reliance on predefined models, offering insights for future tool enhancements.

The remainder of this paper is structured as follows: Section 2 reviews background and related work on AI-powered software testing. Section 3 presents the research design and planning for the study. Section 4 describes the Systematic Tool Review (STR) methodology, while Section 5 presents its results. Section 6 details the empirical study methodology and findings. Section 7 concludes the study and discusses future research directions.

# 2 BACKGROUND AND RELATED WORK

In this section, we provide brief background on emergence of AI-powered testing tools, and then a review of the related work.

## 2.1 Background: Emergence of AI-powered testing tools

The rise of AI in software testing has introduced a new generation of AI-powered testing tools [1-3]. By reviewing online resources, we find out that different terms have been used to refer to software testing tools, in which AI is an enabler. One could format the phrase as: "AI-X software testing tools", where "X" could be any of the followings: "powered", "based", "assisted", "augmented", "infused", "driven" and "enabled".

We provide in Table 1 the list of all the different terms that we found and are used to refer to software testing tools using AI. Although precise usage of these terms does not seem to have been standardized yet, the terms could have implications with respect to the levels of AI involvement (integration) in a given tool.

The above levels of AI involvement in tools also relate to the six levels of autonomy in software testing [7], as shown in Figure 1. This model has been adapted from the Society of Automotive Engineers (SAE)' model for self-driving cars. Figure 1 illustrates the gradual progression toward full automation as follows:

- Level 0 (Manual Testing): Test execution is entirely manual, relying on the tester's skills and domain knowledge. No automation is used.
- Level 1 (Assisted Automation): Some test execution tasks are automated, but a tester remains actively involved in test design, execution, and validation.
- Level 2 (Partial Automation): Testers use codeless automation tools to generate test scripts, but human intervention is still required for test maintenance and decision-making.
- Level 3 (Accelerated Automation / Self-Healing Tests): AI-powered self-healing mechanisms can update test scripts when UI elements change, reducing test maintenance needs. However, human oversight is still needed to validate AI decisions.
- Level 4 (Supervised AI-powered Testing): AI is capable of generating and executing test suites, with minimal human intervention. Testers act as supervisors, reviewing AI's outputs and intervening if required.
- Level 5 (Fully Autonomous Testing): AI completely manages test creation, execution, maintenance, and defect detection without human involvement.

**Table 1-Different terms used to refer to software testing tools using AI. The terms could have implications with respect to AI involvement (integration) levels**

| Term | Level of AI involvement (integration) in the tool | Description |
|---|---|---|
| AI-Assisted | Minimal | AI provides limited support, assisting testers but not taking the lead in testing activities. |
| AI-Infused | Low | AI enhances existing workflows but is not the primary decision-maker in automation. |
| AI-Augmented | Moderate | AI extends and optimizes traditional automation but still relies on human supervision. |
| AI-Enabled | Moderate | AI functionalities exist alongside traditional automation but do not drive major decisions. |
| AI-Based | Moderate | AI is fundamental to the tool, but manual control is still necessary for validation. |
| AI-Driven | High | AI plays a major role, dynamically generating and executing tests with minimal human input. |
| AI-Orchestrated | High | AI actively manages the execution of multiple test processes, optimizing automation pipelines. |
| AI-Automated | Very High | AI is the primary execution agent, handling test automation with little to no manual intervention. |
| AI-Led | Very High | AI is actively making critical testing decisions based on historical data and system behavior. |
| AI-Powered | Full | AI is the core foundation of the tool, making autonomous decisions in testing workflows. |
| AI-First | Full | AI is central to the testing strategy, with all testing processes designed around AI capabilities from the outset. |
| AI-Autonomous | Theoretical / Not Yet Feasible | AI fully manages the entire test lifecycle without human intervention. |

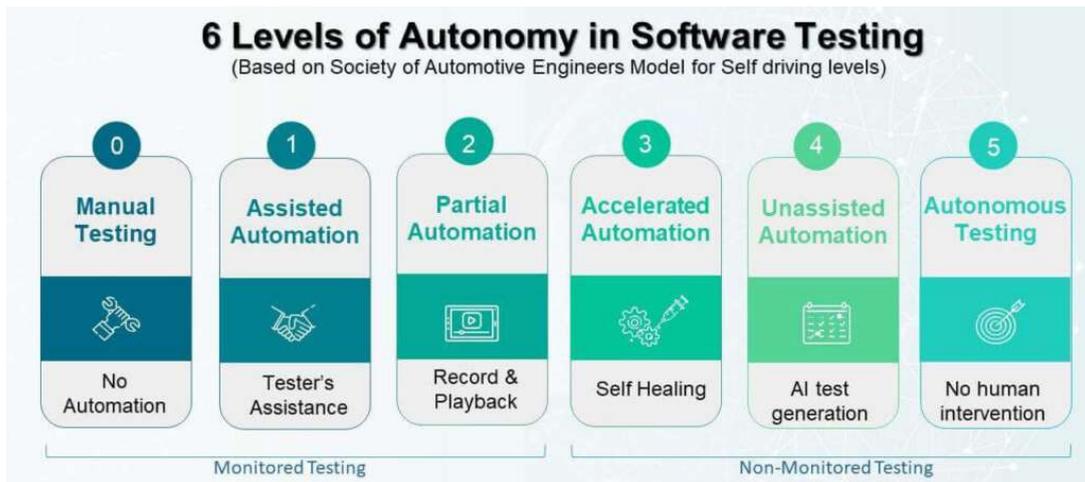

When discussing AI-powered testing tools, it is important to distinguish general-purpose AI tools such as OpenAI ChatGPT and GitHub Copilot, and specific-purpose AI-powered tools exclusively built for software testing, such as tools that we review in this paper (see Section 5.1).

General-purpose AI tools are not built for software testing only, and instead are either broadly general in purpose, such as case of ChatGPT which can be used for any purpose, or are software-engineering focused AI tools, such as GitHub Copilot.

Although not built for software testing exclusively, general-purpose AI tools can provide various types of assistance to test engineers in different software test activities [1-3], e.g., generating draft test plan documents, generating test cases for manual testing, generating test automation code (e.g., in JUnit), and analyzing test log and defect files. Accuracy of outputs generated by general-purpose AI tools for different test activities can vary and is dependent on a variety of factors, mainly the amount of information provided by the human testers to the AI tools, and the precision of the "prompt[s" (commands) given by the human tester to the AI tool, an important emerging topic which is called Prompt Engineering [8].

On the other hand, specific-purpose AI-powered tools are dedicated narrow-focus tools purposefully-built to support a selected few test-support features. For example, Parasoft Selenic (parasoft.com/products/parasoft-selenic) is one such commercial tool, and its main feature is AI-powered self-healing of UI test-code [9]. This feature refers to the ability of testing tools to automatically detect and fix broken locators (i.e., references to UI elements like buttons, text fields, or dropdown menus) in test scripts (mostly in the Selenium test framework) without requiring manual intervention from software testers.

## 2.2 Related work

Systematic reviews of various types have been widely published in software engineering, e.g., Systematic Literature Reviews (SLRs) [10] and Multivocal Literature Reviews (MLRs) [11].

Several prior review papers (secondary studies) have explored the area of AI-powered testing and the tools in this area. Table 2 lists the secondary studies in this area, and a summary of their contributions. While these related review papers provide valuable insights, our work in comparison can be considered the most in-depth study on the existing AI-powered software testing tools, as we have systematically screened and reviewed all the tools in this area, found by the end of 2024 (56 tools). Furthermore, we review the tools' features, benefits and limitations via three research questions (RQs), as briefly listed in Table 2 and discussed in Section 3. No prior work has conducted such an analysis at this level of depth.

Table 2-Related secondary studies in this area

| Paper reference | Year | Title | Number of papers reviewed | Number of tools reviewed | Focus / contributions of the paper |
|---|---|---|---|---|---|
|  |  |  |  |  |  |

| [12] | 2017 | AI automation and it's future in the United States | 22 | 3 | The paper focus is on the role of AI automation in the United States. It reviews the impacts of AI automation in general. AI automation in software testing is only a small part of the article. |
| [5] | 2021 | AI-powered software testing: A grey literature analysis | 40 | 6 | Mentions the features the tools offer and how they solve challenges in test automation. Focuses on problems faced by practitioners in test automation. |
| [6] | 2021 | A cognitive approach in software automation testing | 13 | 1 | The paper mentions the algorithms used in AI features provided by testing tools |
| [13] | 2023 | A review of AI-augmented end-to-end test automation tools | 18 | 8 | Identifies primary AI techniques used in test tools |
| - | 2025 | This work | - | 56 | Three RQs: <ul><li>RQ1: Classifying AI features of the tools</li><li>RQ2: Reviewing the benefits of the AI-powered features</li><li>RQ3: Limitations of the AI-powered features</li></ul> |

## 3 RESEARCH METHOD AND APPROACH

This study follows a two-phase research approach to investigate AI-powered testing tools:

1. A Systematic Tool Review (STR) to categorize and analyze AI-powered testing tools
2. A follow-up empirical study to evaluate the benefits and limitations of two selected AI-powered testing tools for testing two selected software under test (SUT)

The design and execution details of the STR are presented in Section 4, while its results are in Section 5. The empirical study methodology and findings are described in Section 6.

This study is guided by the following research questions (RQs):

- RQ1: What AI-powered testing tools exist in the industry and what are their features?
- RQ2: What benefits can the AI-powered features of the tools provide to test engineers, in increasing effectiveness and efficiency of software testing?
- RQ3: What are the limitations of existing AI-powered testing tools, which could serve as insights for future tool enhancements and further research?

RQ1 is explored in the STR, while RQ2 and RQ3 are explored in both the STR and the empirical study. Each RQs informs a different aspect of the study, ensuring a comprehensive evaluation of AI-powered software testing. This structured approach enables both a broad landscape analysis (STR) and practical validation (empirical study), contributing to a thorough understanding of AI-powered software testing.

## 4 DESIGN AND EXECUTION OF THE SYSTEMATIC TOOL REVIEW

We designed and conducted the Systematic Tool Review (STR) using established guidelines for systematic reviews in software engineering [10, 11] to ensure a structured and objective assessment of the tools in this area. We discuss in the rest of this section the details of the STR process.

### 4.1 Approach for searching for tools

We performed the search for tools in the Google search engine, since our goal was to find industry tools, and not those which may have been proposed only in research papers.

Our search string format was: "AI-X software testing tools", where "X" was replaced with different terms used in industry to refer to these tools (as discussed in Section 2.1): "powered", "based", "assisted", "augmented", "infused", "driven" and "enabled".

The above Google search process yielded 74 tools, which we then screened by applying a set of inclusion and exclusion criteria, as discussed next.

## 4.2 Inclusion and exclusion criteria

The inclusion criteria for selecting related AI-powered testing tools in our study were:

- Tools explicitly using AI: Only tools whose product sheets (web pages) explicitly mentioned using AI were included.
- Publicly available (open-source or commercial with accessible documentation). We are aware that certain companies have developed and use related (internal) tools / technologies, but they have not made their tools public. Obviously, we have no way of knowing about such tools.

The exclusion criteria were:

- General-purpose automation tools without AI-powered features
- Proprietary tools with no publicly available information

After application of the inclusion and exclusion criteria, 56 AI-powered testing tools remained for analysis.

## 4.3 Approach for extraction, management and synthesis of data

We discuss next our approach for data extraction, management of data and also for synthesis of findings [14, 15].

### 4.3.1 Data extraction approach for each RQ

For each tool, data extraction was performed to address the study's three RQs. This process systematically analyzed official tool documentation, demonstration videos, and user feedback, ensuring an objective and structured evaluation.

- RQ1 – AI Features: We extracted the list of AI-powered features for each tool by carefully reviewing its homepage, and demo videos.
- RQ2 & RQ3 – Effectiveness, Efficiency, and Limitations: Due to time constraints, we did not install or try each tool. Instead, our assessment relied on an objective evaluation of the following three data sources:
  - The features and functionality described in tool documentation and online demos.
  - User reviews from www.g2.com, which provided practitioner insights into tool effectiveness, efficiency, and limitations. g2.com is a widely recognized platform for reviews on software applications. The g2.com website states that: *"Find the right software and services based on 2,917,200+ real reviews".* For most of the tools under review, we found extensive feedback (reviews) from tool users (test engineers using those tools) discussing benefits and limitations of each tool. These reviews provided practitioner insights into a given tool's usability, effectiveness, and limitations when applied in real-world large-scale testing projects.
  - In certain cases, when from the above available data sources, we were not able to extract certain features or data about a given test tool, we interacted with the ChatGPT LLM tool to ask questions about the test tool. In Appendix B, we present a few examples of those interactions with the ChatGPT tool.

The above approach allowed for a systematic yet practical analysis of AI-powered testing tools while ensuring that insights were derived from multiple reliable sources.

### 4.3.2 Data management: Google spreadsheet

All extracted data was systematically recorded in an online Google spreadsheet, serving as a structured repository for tool categorization and analysis. The online dataset is available at bit.ly/AI_testing_tools, and contains:

- The list of 56 AI-powered testing tools, and the categorization of their AI-powered features
- A systematic mapping (traceability) of tools to the RQs, ensuring alignment between extracted data and study objectives
- Testers' reviews (user feedback) from g2.com, who have actually used the tools. From those user feedback data, we extracted the benefits and limitations of AI features of each tool

### 4.3.3 Approach for synthesis of qualitative data and evidence

As discussed above, a subset of the extracted data in our tool review was qualitative data in forms of descriptive text, e.g., benefits and limitations of AI features of a given tool, either as user feedback or descriptive data from tool documentation.

For synthesizing findings out of those qualitative data, we used the qualitative coding approach, which is established qualitative analysis and grounded theory approach in software engineering [14, 15]. Further details will be provided in Sections 5.2 and 5.3, in which we present the findings regarding those two issues (corresponding to RQ2 and RQ3 of the study).

# 5 RESULTS OF THE SYSTEMATIC TOOL REVIEW

This section presents the findings from the Systematic Tool Review (STR), analyzing AI-powered testing tools based on the three research questions (RQs). The extracted data stored in the Google spreadsheet serves as the basis for the reported results.

## 5.1 RQ1: List of the AI-powered testing tools and classifications of their AI features

We provide in Table 3 the list of the tools and classifications of their AI features, that were found and reviewed via the systematic process in our work. Table 2 also shows the number of tools supporting each AI-powered feature, serving as an indicator of industry focus areas and market demand for AI-powered testing capabilities.

For brevity, we do not include the URLs for each tool, but they can be easily found through internet search. Below, we provide an in-depth review of each AI-powered feature category.

### 5.1.1 AI-powered test generation

43 tools (77% of the tools under review) provide AI-powered test generation. This feature refers to the use of AI to automatically create test cases and/or test-automation code.

While traditionally focused on analyzing application behavior and historical defects to generate test cases, modern AI-powered tools also leverage additional input sources and approaches. These include code analysis (identifying high-risk areas through static or dynamic analysis), requirements and specifications (using Natural Language Processing to extract test scenarios from documentation), user behavior analytics (generating tests based on real user interactions), test coverage gaps (targeting under-tested code areas), and configuration data (creating tests for different environments, devices, and system settings).

By integrating these diverse data sources, AI-powered testing tools can provide large test suites for a given System Under Test (SUT) in a few seconds or minutes, a task which could have taken an experienced software test engineer hours or days.

In our review of the tools, we found out that the tools have taken different approaches to AI-powered test generation:

1. low (or no) code approach for test generation using AI-powered NLP (20 tools)
2. low (or no) code approach using record / playback (9 tools)
3. AI-powered analysis of code-coverage (13 tools)
4. other approaches such as AI-powered model-based testing (3 tools)

We discuss each of the above four AI-powered test generation approaches next.

#### Test generation using AI-powered NLP

This feature leverages AI to enable non-technical users to create automated test cases using natural languages instead of traditional programming syntax. This technology allows users, such as business analysts or product owners, to write test scenarios in plain English (or other natural languages), which the AI-powered system then interprets and converts into executable test scripts automatically. This approach reduces the technical barrier to creating automated tests, allowing broader team collaboration in the software development lifecycle.

For example, a user could write, *"Verify that the login button redirects the user to the dashboard after entering valid credentials"* and the AI-powered tool would translate this into a structured script compatible with automation frameworks such as Selenium or Appium.

This not only speeds up test development but also ensures better alignment between business requirements and testing outcomes. Moreover, it minimizes human errors, enhances collaboration, and streamlines the automation process.

A concrete example from the above approach is shown in Figure 2. This example is from an insightful demo video of the ACCELQ AI-powered test tool (link is shown in the figure caption).

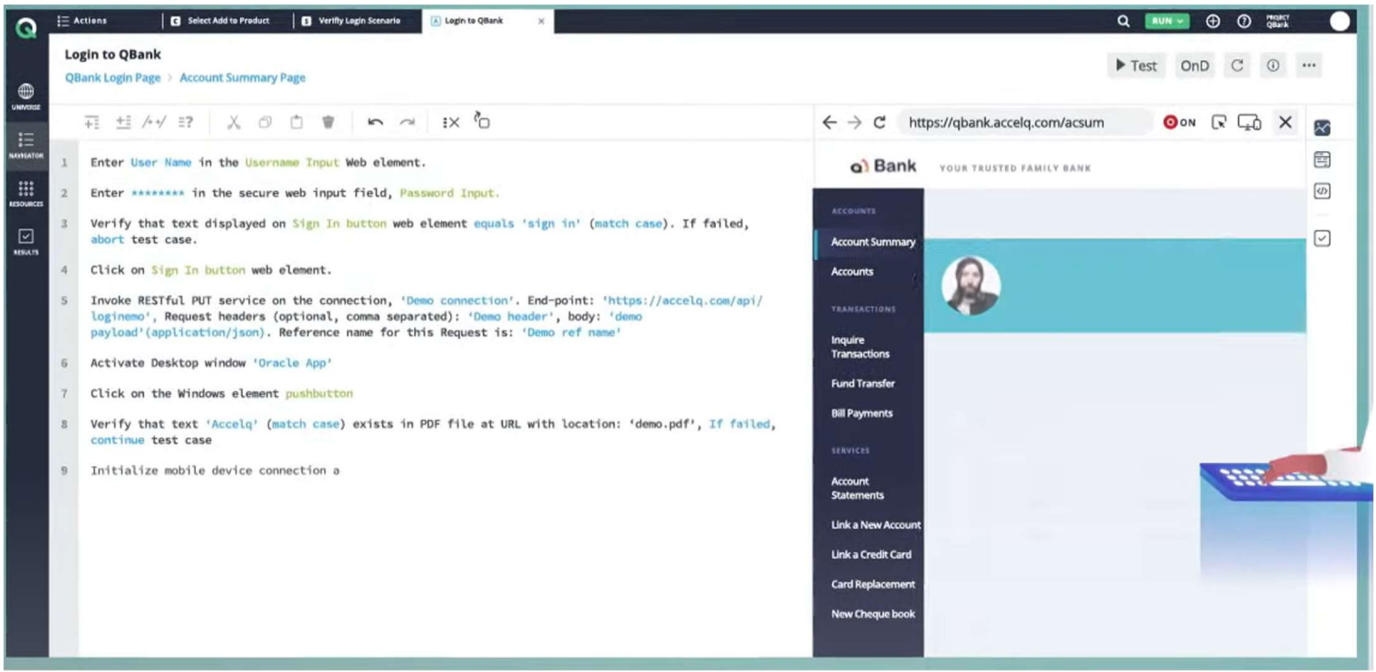

Figure 2- An example of the AI-powered NLP-based no-code approach for test generation (video source: bit.ly/ACCELQ-AI-codeless-testing).

## Test generation using AI-powered test recording

AI-powered test recording simplifies test-case design and development by enabling record-and-replay functionality with AI-enhanced adaptability. This feature is supported by 9 out of 56 tools (16%) in the set of tool.

Traditional record-and-playback tools allow testers to capture user interactions and replay them to automate test execution. However, these methods have historically struggled with test flakiness, where recorded scripts break due to UI changes, timing issues, or environmental variations. AI-powered test recording enhances this approach by improving element recognition, timing synchronization, and automated test adaptation. AI-powered test recording tools introduce several further improvements over traditional record-and-playback automation, including:

- Smart Wait Handling – AI detects dynamic UI loading patterns and automatically adjusts test execution timing to avoid false failures caused by delayed element rendering (tools such as: Functionize, Testsigma).
- Adaptive Test Generation – Some tools use machine learning models to analyze recorded test sessions and suggest additional test cases based on historical defects or risk-based prioritization (tools such as: Virtuoso, TestRigor).
- Automated Data Parameterization – AI dynamically extracts input values from recorded interactions, allowing test scripts to be automatically converted into reusable, data-driven tests (tools such as: ACCELQ, Testim).

Taken from a demo video of one of the tools (named CloudQA), we provide an example of AI-powered no-code test recording in Figure 3. In this screenshot, we can see that AI-powered test recording enables testers to interact with the SUT, while the AI automatically:

- Captures clicks, inputs, and navigation flows
- Converts recorded interactions into structured test scripts
- Detects pattern-based variations in UI elements to ensure test robustness
- Provides visual test editing, allowing testers to modify test steps without coding

This approach is particularly beneficial for non-technical testers who may lack scripting expertise but need to create reliable and adaptable automated tests.

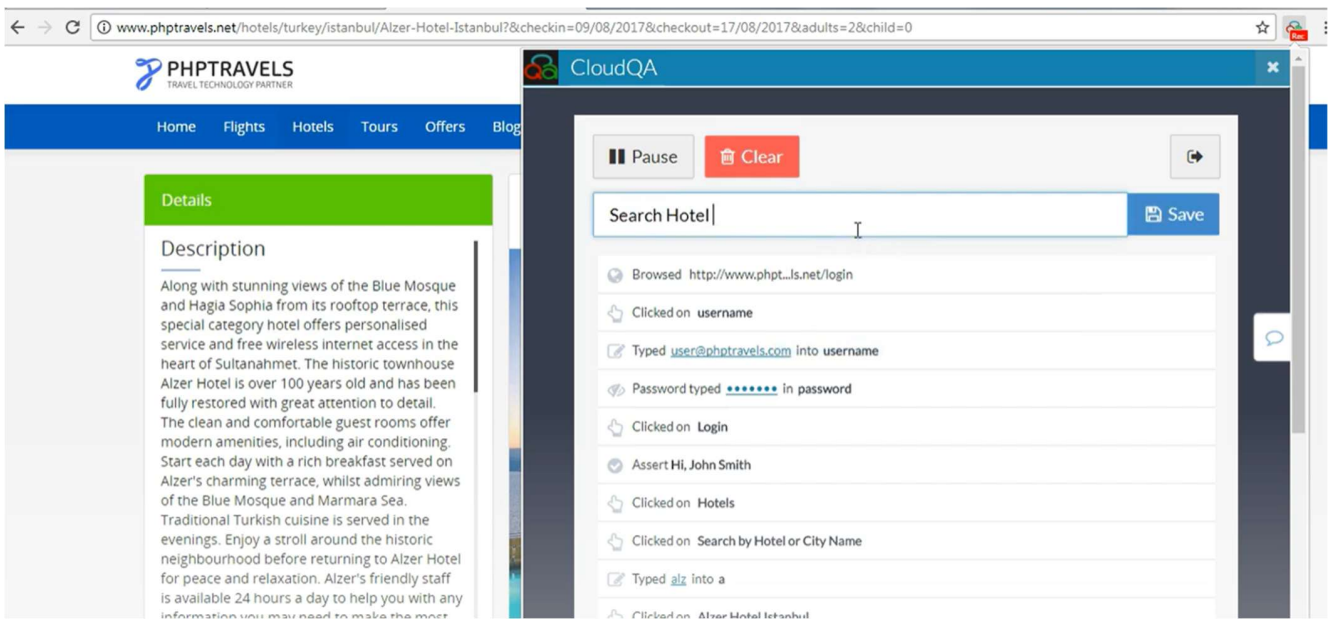

Figure 3- AI-powered test recording in action, showing how AI automatically captures user interactions and generates test scripts in real-time (video source: bit.ly/CloudQA-AI-codeless-testing).

### Test generation using AI-powered coverage analysis

Test generation using AI-powered code-coverage analysis is another widely adopted feature that identifies untested or high-risk areas in code (e.g., modules with high cyclomatic complexity) and automatically generates targeted test cases to improve coverage (offered by tools such as Diffblue Cover and TestRigor). Some tools claim to achieve high coverage with minimal human intervention; for instance, the website of the BaseRock tool states that it can provide *"up to 80% code coverage without developer intervention"*. This capability can help organizations reduce the risk of undetected defects by ensuring that critical system components receive sufficient test coverage.

AI-powered tools analyze code structure, control flow, and defect history, using static and dynamic analysis to detect untested regions and vulnerable code paths. By leveraging machine learning models trained on historical defect patterns, AI can predict high-risk areas and automatically prioritize test generation.

### Test generation using other AI-powered approaches

Beyond the above AI-powered features offered by many tools, some tools offer other AI-powered features to generate test cases. These methods leverage advanced AI techniques, such as model-based testing, agentic AI-powered workflows, and behavioral analytics, to create more context-aware and adaptive test cases.

In tools offering AI-powered model-based testing, AI constructs system models based on application behavior, architecture, or requirements. These models allow the AI to autonomously generate test cases that reflect various system states, ensuring better coverage of edge cases and system workflows. This technique is particularly beneficial for complex software systems with numerous interactions, where traditional script-based automation may fail to capture all dependencies. Two example tools in this category are Keysight Eggplant and Functionize.

Another emerging technique involves agentic AI-powered workflows, where AI models actively guide test creation and execution. Instead of passively generating test cases, the AI continuously interacts with the system under test (SUT), learning from its responses and refining its testing strategy in real time. Some tools utilize chat-based AI assistants to assist testers in dynamically refining and extending their test cases based on evolving software requirements (tools such as: Qodo, Testim).

Additionally, AI-powered test generation can leverage user behavior analytics, where real-world usage patterns are analyzed to generate test cases that mimic actual user interactions. By studying historical user data, logs, and telemetry, AI can prioritize high-impact test scenarios, ensuring that automation aligns with real-world usage conditions. This data-driven approach enhances the relevance and effectiveness of automated tests, particularly in web and mobile applications that have diverse user bases and varying interaction patterns. Two example tools in this category are LoadMill and Testify Pulse.

### 5.1.2 AI-powered self-healing of UI test-code

This feature refers to the ability of testing tools to automatically detect and fix broken UI locators in test scripts without requiring manual intervention from software testers. UI locators are references to UI elements like buttons, text fields, or dropdown menus in the UI of a SUT, in test frameworks such as Selenium.

32 tools (57% of the tools under review) provide AI-powered self-healing of UI test-code. This feature enables automated adaptation (repair) of broken test scripts when UI elements change, reducing test maintenance costs for test engineers and improving test stability.

In conventional automated testing, scripts rely on predefined locators such as IDs, XPaths, or CSS selectors to interact with UI elements. However, when the front-end code or design of the application changes (e.g., a developer renames a button's ID), these locators can break, causing test failures. Without self-healing capabilities, a tester would need to manually update each broken locator, a time-consuming and error-prone process. Self-healing tools powered by AI and machine learning can automatically recognize when an element locator breaks and intelligently find alternative locators by analyzing patterns such as the element's attributes, its hierarchy in HTML DOM (Document Object Model), or visual similarity. The AI tool can then update the script automatically, reducing maintenance efforts and minimizing disruptions in the testing process.

For example, consider an automated test script that clicks a "Submit" button on an HTML page, defined as: `<button id="submit-btn">Submit</button>`. The button was initially identified using the following Java Selenium code snippet, using the find-by ID locator:

```
// Locate the button using its original ID and click it
WebElement submitButton = driver.findElement(By.id("submit-btn"));
submitButton.click();
```

Later, a front-end developer of the SUT changes the button's ID to: `<button id="send-btn">Submit</button>`. In traditional test automation, this change would cause the test script to fail because it cannot find the element with `id="submit-btn"` anymore. With self-healing automation, the AI tool can detect that the original locator no longer works but can identify the button by other attributes, such as its text content ("Submit") or position on the page. It automatically updates the script with the new locator or adapts the strategy (e.g., switching from an ID-based locator to a text-based one), allowing the test to continue running without testers' intervention.

### 5.1.3 AI-powered visual testing

AI-powered visual testing enhances software quality assurance by detecting UI inconsistencies, layout shifts, and visual regressions across different environments. This feature is implemented in 16 out of 56 AI-powered testing tools (29%). Traditional visual testing relies on pixel-to-pixel comparisons, which often produce false positives due to minor rendering differences. AI-powered approaches leverage computer vision and machine learning algorithms to intelligently analyze UI changes, distinguishing intentional modifications from unintended visual regressions with higher accuracy.

These regressions could be unexpected changes or errors in how the UI appears when viewed on different screen resolutions (e.g., mobile, tablet, or desktop) or in various browsers (e.g., Chrome, Firefox, Safari). For example, if a button is misaligned on a mobile screen or an image does not display properly on Firefox, AI can automatically detect these issues during testing without requiring manual inspection by a tester. By analyzing both pixel variations and structural changes in the DOM, AI-powered tools improve test reliability by focusing on meaningful visual defects while reducing false positives (tools such as: Applitools, SmartBear VisualTest).

In these tools, AI models compare screenshots captured from previous test runs against updated versions, highlighting differences and allowing test engineers to approve or reject changes. This automation significantly reduces manual verification effort, making visual testing scalable for large applications with frequent UI updates.

In summary, AI-powered visual testing enhances traditional UI validation by leveraging computer vision and machine learning to detect visual defects more intelligently and reduce false positives. While it provides significant efficiency gains and integrates well with automation workflows, it still requires human oversight to refine AI-generated results and address challenges related to dynamic content and false positives.

### 5.1.4 AI-powered test-result and failure analysis

11 tools (20% of the tools under review) provide AI-powered test-result and failure analysis features. These features automate root cause identification, reducing debugging time for testers. These tools leverage machine learning and pattern recognition to analyze logs, execution results, and historical defect data. By detecting recurring failure patterns, AI can suggest probable causes, categorize failures (e.g., flaky tests vs. real defects), and recommend fixes (tools such as Appsurify TestBrain, and Copado Robotic Testing).

Some tools apply AI-powered clustering to group similar failures, helping testers prioritize issues efficiently (tools such as Testify Pulse, and ReportPortal). Others integrate with CI/CD pipelines to provide real-time diagnostics, reducing release delays (tools such as Tricentis Tosca, and LoadMill). AI can also analyze operational logs to detect performance bottlenecks and API failures (tools such as Parasoft SOAtest, and TestResults).

Another feature under this category are leveraging AI models to analyze large volumes of test data, identifying unusual patterns or deviations from expected behavior that could signal defects or performance issues. By continuously monitoring test results over time, these models learn to detect anomalies—such as unexpected spikes in response times or sudden increases in test failures—that may indicate underlying issues in the software.

In addition to detecting anomalies, predictive analytics use historical data to forecast defect-prone areas within the codebase, enabling testers and developers to focus their efforts on high-risk components. This targeted approach helps optimize test execution by prioritizing critical test cases and reducing unnecessary testing on stable code sections. For instance, if a particular module frequently experiences issues after updates, the AI system can flag it for more intensive testing in future cycles. This proactive strategy improves software quality, reduces testing time, and enhances resource allocation across the development process

### 5.1.5 Other AI-powered features

Beyond the four categories of AI-powered features discussed above, 23 tools in the reviewed set provide additional AI-powered features, enhancing various aspects of software testing. These capabilities extend the role of AI beyond test creation and maintenance, supporting test selection, failure prediction, exploratory testing, and test data generation.

One important capability is AI-powered test prioritization, which enables tools to identify and focus testing efforts on high-risk areas based on historical defect data, code complexity, and recent code changes. By analyzing these factors, AI can intelligently order test execution, ensuring that critical tests run first, reducing unnecessary test execution and optimizing CI/CD pipelines (tools such as: Avo Automation, Testify Pulse).

Another widely adopted feature is AI-powered predictive analytics, which anticipate potential failures before they occur. By examining historical test data, defect trends, and system behavior patterns, AI can identify code segments likely to introduce defects and recommend preemptive testing. This approach improves early defect detection, allowing teams to address issues before they impact production (tools such as: Copado Robotic Testing, LoadMill).

AI-powered exploratory testing enhances traditional testing by automatically navigating applications, interacting with UI elements, and detecting anomalies without predefined scripts. This technique allows AI to simulate real user behavior, uncovering defects that scripted tests might miss. It is particularly useful for testing new features, evolving applications, and non-deterministic systems, where predefined test scripts are difficult to maintain (tools such as: Keysight Eggplant Test).

Another valuable AI-powered capability is automated test-data generation, which can be used to create realistic and diverse test inputs. AI-powered test-data generation ensures comprehensive coverage of edge cases, reducing the effort required for manual test data preparation. By analyzing patterns in production data (usage logs), AI can generate synthetic test data that closely resembles real-world usage scenarios, improving the accuracy of test cases. This feature is particularly useful in data-driven testing, API testing, and security validation, where varied and high-quality test data is essential (tools such as: Virtuoso, Avo Automation).

Table 3 –The AI testing tools and classification of their features

| Tool names | AI features of the tools | | | | | If test generation, what approach is used? | | | | Level (type) of tests generated / repaired | | | |
|---|---|---|---|---|---|---|---|---|---|---|---|---|---|
| | Test generation | Self-healing of UI tests | Visual testing | Test-result and failure analysis | Other AI features | Low (no) Code, using NLP | Low (no) Code, using Record / Playback | Code-coverage analysis | Other | Unit tests | API (integration) tests | UI (system) tests | Only test cases, no test code |
| # of tools under each category → | 43 | 32 | 16 | 11 | 23 | 20 | 9 | 13 | 3 | 16 | 3 | 20 | 2 |
| ACCELQ | x | x | | | | x | | | | | | x | |
| Amazon CodeWhisperer | x | | | | | x | | | | x | | | |
| Applitools | | x | x | | AI is used to ignore dynamic content like ads, reducing unnecessary tests | | | | | | | | |
| Appsurify - TestBrain | | x | | | Uses AI to provide alerts on risky code changes to catch defects early. | | | | | | | | |
| Appvance | x | x | | | | | | x | | | | x | |
| Aqua | x | | | | Test prioritization | | | | | | | x | |
| Autify | x | x | x | | | | | | | | | x | |
| Avo Automation | x | x | | | • Change-impact analysis • Automated test-data generation | x | | | | x | | | |
| BaseRock (formerly: Sapient) | x | x | | | Creating realistic synthetic test data | | | x | | x | x | | |
| BrowserStack - Percy | | | x | | | | | | | | | | |
| Cloud QA | x | x | | | | | x | x | | | | x | |
| Code Test Generator | x | | | | | | | x | | x | | | |
| qodo.ai (formerly: Codium.io) | x | x | | | • Agentic test workflows: Chat-based, guided test generation for tailored testing • AI-powered test suite extension and validation • Context-Aware test-code customization | | | x | | x | x | | |
| CommandDash.io (formerly: welltested.ai) | x | | | | | | | x | | x | | | |
| Copado - Robotic Testing | x | x | | | Predictive analytics for detailed reporting | x | | | | | | x | |
| diffblue Cover | x | | | | Provides feedback on testability of code | | | x | | x | | | |
| digital.ai | x | x | | | | x | x | | | | | x | |
| Functionize | x | x | x | | A gen-AI feature named testASSIST enables non-technical users to manipulate testing scenarios and streamline their test case creation process | x | | | | | | x | |
| Gemini Code Assist (formerly Duet AI) - Google | x | | | | | | | x | | x | | | |
| Github CoPilot | x | | | | | x | | x | | x | | x | |
| Healenium | | x | | | | | | | | | | | |
| iHarmony | x | x | | | | | x | | | | | | |
| IntelliJ IDEA | x | | | | | | | | | x | | | |
| Katalon | x | x | x | x | | x | | | | | | web apps | |
| Keysight - Eggplant Test | x | | x | | AI-powered exploratory testing | x | | | Model-based testing | | | x | |
| Kobiton | x | x | x | | | | | x | | | | mobile apps | |
| LambdaTest HyperExecute | | | x | | An AI feature named Test Orchestration: automatically groups and distributes tests intelligently across different testing environments | | | | | | | | |
| Launchable | | | x | | | | | | | | | | |
| LoadMill | x | x | | x | | | | | User behavior (operational profile, logs) | | | x | |

|  | | | | |  | | | | | | |
|---|---|---|---|---|---|---|---|---|---|---|---|
| Mabl | x | x | | | | x | | | | | |
| Machinet AI Unit Tests (a plugin for IntelliJ) | x | | | | --AI describes what each test method is doing<br>--AI generates mocks, objects, and non-trivial assert statements<br>--AI decides whether to use mocks or not. | | | x | x | | |
| Parasoft JTest | x | | | | | | | x | x | | |
| Parasoft Selenic | | x | | x | | | | | | | |
| Parasoft SOAtest | x | | | | | x | | | API logs (recorded traffic) | x | | |
| pCloudy | x | x | x | | Observability agent, for full testing visibility | x | | | | x | |
| Perfecto | | x | | | | | | | | | |
| ReportPortal | | | | x | | | | | | | |
| retest | x | | x | | | | x | | | x | |
| SauceLabs | x | x | x | x | | x | | | | x | |
| Smartbear - VisualTest | | | x | | AI-powered object recognition | | | | | | |
| Smartbear - TestComplete | | x | | | | | | | | | |
| Sofy Co-Pilot | x | | x | x | --Hyper-intelligent (optimized) regression testing<br>--Possibility of further training of the built-in AI model | x | | | | mobile apps | |
| taskade | x | | | | | | | | | | x |
| Tabnine | x | | | | | | | | x | | |
| Test Grid | x | x | | x | | x | | | | | |
| Testify - Pulse | x | | | x | --AI-hierarchical clustering: groups API behaviors based on similarities, identifying both documented and undocumented patterns. It ensures each pattern is tested, improving coverage and detecting edge cases.<br>--AI dependency analysis: Maps system component relationships for better test generation | | | x | x | | |
| TestingWhiz | x | x | x | | | | x | | | | |
| TestResults | | x | x | | --AI Virtual user tests the software as a user would.<br>--Visual hints: AI-powered spatial algorithm to detect interaction points without predefined hints, enabling adaptive and robust automated testing. | | | | | | |
| TestRigor | x | x | | | | x | | | | | |
| TestSigma | | x | | | | x | | | | | |
| Tricentis Testim | x | x | | | --Learns user flows, recognizes repeated sequences and suggests reusable elements.<br>--Captures screenshot of every test step and compares to previous screenshots to identify what has been changed, to suggest appropriate test cases.<br>--AI helps in the development of well-architected, clean tests that optimize reuse and minimize maintenance. | | x | | | x | |
| Tricentis Tosca | x | x | x | | --A feature named RiskAI analyzes changes in SAP systems to identify high-risk areas needing prioritized testing, reducing release scope while achieving risk coverage | x | x | | | | x |
| Unit-test.dev | x | | | | | | | | x | | |
| Virtuoso | x | x | | | Test data generation | x | | | | web apps | |
| Webo.ai (by Webomates) | x | x | | | | | x | | | web apps | |
| Workik | x | | | | | | | x | x | | |

## 5.2 RQ2: Benefits of AI features in increasing test effectiveness and efficiency

The main reason for the development and inclusion of AI features in the tools is to support test engineers and enhance their effectiveness and efficiency in software testing. AI-powered capabilities aim to reduce manual effort, improve defect detection, and optimize test execution, ultimately helping teams deliver higher-quality software with reduced time-to-market.

As discussed in Section 4.3, to assess the benefits of AI features, one of our data sources was www.g2.com, a widely recognized platform for software reviews. In addition to G2.com reviews, we analyzed official tool documentation, online vendor reports, product demonstrations, and industry white papers. These sources provided further insights into how vendors position their AI-powered features, the specific benefits they claim, and how these tools integrate into existing test automation pipelines. Moreover, we reviewed a large number of online demo videos, allowing us to observe how these tools function in practice and verify whether vendor claims aligned with practitioner experiences.

Once all qualitative (descriptive) data was collected, we applied the qualitative coding technique [14, 15] to systematically cluster, synthesize, and extract key themes regarding the benefits of AI-powered testing features. This approach ensured a structured analysis, allowing us to identify common patterns across multiple data sources. However, as discussed in Section 4.3.1, we shall note that, due to our resource and time constraints, we did not conduct our own empirical studies of each tool to directly measure the effectiveness and efficiency of its AI features. Instead, our findings in the systematic tool review are based on secondary data analysis, reflecting both vendor claims and real-world user experiences. However, as we will report in Section 6 of this paper, we empirically assessed two of the tools ourselves.

### 5.2.1 Potential increase in test effectiveness using AI features

AI-powered testing tools can enhance test effectiveness by improving defect detection rates, expanding test coverage, and reducing human bias in test design and test selection. Unlike test efficiency, which focuses on reducing effort and time, test effectiveness concerns how well testing detects defects, ensures software reliability, and increases confidence in the SUT.

The findings indicate that AI-generated test artifacts could contribute to improving test effectiveness in several ways, as discussed next.

#### 1. More comprehensive test coverage (reducing testing gaps)

One of the main challenges in software testing is identifying untested scenarios and missing test cases. AI-generated test artifacts can expand coverage by analyzing code structure, system behavior, and historical defect data, identifying potential test gaps that manual testers might overlook. This is particularly useful in large-scale and complex applications, where traditional test design may fail to cover all execution paths. AI-powered tools can use machine learning models to predict areas prone to defects and ensure that under-tested code regions receive adequate test coverage (example tools: Diffblue Cover, TestRigor).

#### 2. Higher defect detection rate (catching more bugs)

Traditional testing often relies on predefined test suites, which may fail to capture unexpected failure patterns. AI-powered tools can potentially improve defect detection by dynamically analyzing runtime behavior, execution logs, and edge cases that manual or script-based testing might miss. AI can detect flaky tests, flag patterns in software behavior that indicate potential failures, and identify anomalies in test execution results, making it more effective in catching hidden defects and regressions. This can particularly be valuable in highly dynamic applications, such as web and mobile platforms, where UI elements and backend responses frequently change (example tools: Applitools, Mabl).

#### 3. Making test coverage more objective (reducing human bias)

Manual test selection often involves subjective decisions, where testers prioritize familiar or frequently used test cases while unintentionally neglecting risk-prone areas. AI-powered testing can eliminate this bias by objectively analyzing risk factors such as historical defect trends, execution logs, and real-world usage data. AI can ensure that test cases align with actual risk levels rather than being based on human intuition, resulting in a more structured and data-driven approach to test coverage (example tools: Functionize, Tricentis Tosca).

#### 4. Prioritizing critical and risky areas of the system

Not all parts of a software system pose the same level of risk. As it is widely known, the 80/20 rule, also known as the Pareto Principle, in software testing suggests that focusing on a small portion (20%) of test cases or code can reveal a significant majority (80%) of defects.

AI-powered testing can dynamically prioritize high-risk areas, ensuring that business-critical and frequently modified components receive more extensive testing than stable, low-risk sections. By leveraging historical defect data, code complexity metrics, and version control changes, AI can determine which parts of the system require more intensive test execution. This approach enhances the overall effectiveness of test execution, reducing the likelihood of defects escaping into production (example tools: Qodo, Testim).

## 5. Detecting defects earlier in the development cycle

Detecting defects early in the development process significantly reduces the cost and effort required for bug fixes. AI-powered testing can enable real-time analysis of code changes and pull requests, automatically generating targeted test cases to catch defects as soon as code is committed. This approach minimizes defect leakage into later stages of the software development lifecycle (SDLC) and reduces the need for extensive post-release testing.

AI can also automate regression testing after each commit, ensuring that newly introduced changes do not cause unexpected failures (example tools: LoadMill, ReportPortal).

## A critical consideration when assessing increase in test effectiveness using AI features: AI can also provide incorrect outputs, thus the need for human oversight (peer review) of AI-generated test artifacts

Despite all its advancements, AI-powered testing is not fully autonomous. AI-generated test artifacts must always undergo human peer review and validation to ensure their correctness, contextual relevance, and practical applicability. This issue is broad and applies to every area which uses AI (i.e., any subject matter in the world has been impacted by and uses AI). Human oversight in AI-generated materials is crucial to ensure accuracy, ethical alignment, and quality, mitigating risks like bias, misinformation, and plagiarism, and fostering trust in AI-generated content. Even the major AI Act of the European Union (EU) has a dedicated article on human oversight [16].

While AI-powered tools enhance test effectiveness by expanding test coverage, identifying high-risk areas, and reducing human bias, they do not possess deep domain understanding and may generate test cases that lack business relevance or fail to capture edge cases.

AI models often rely on historical defect patterns, execution logs, and code complexity metrics to drive test selection, but they cannot fully grasp business logic intricacies, user workflows, or regulatory constraints. This can lead to overgeneration of redundant test cases, ineffective test assertions, or missed defects in complex scenarios. Therefore, test engineers play a crucial role in reviewing AI-generated test artifacts, refining them for real-world applicability, and ensuring that AI-powered optimizations align with actual software risks and business needs.

Additionally, AI-powered defect detection, while promising, can suffer from false positives and misclassifications, requiring human testers to interpret AI-flagged defects and differentiate between critical failures and non-issues. This is particularly important in AI-powered visual testing, where AI may incorrectly identify minor UI shifts as regressions or fail to highlight functional issues embedded within a visually unchanged interface.

Ultimately, AI should be seen as an augmentation of human testers rather than a replacement. By collaborating with AI, test engineers can maximize AI's efficiency while applying their judgment to refine test coverage, validate failures, and ensure high-quality software testing outcomes. This hybrid approach allows organizations to leverage AI's automation strengths while maintaining the flexibility and contextual intelligence of human decision-making.

### 5.2.2 Potential increase in test efficiency using AI features

AI-powered testing tools can enhance test efficiency by reducing manual effort, accelerating test execution, and optimizing resource usage. Unlike test effectiveness, which focuses on improving defect detection and test quality, test efficiency improvements relate to reducing the time, effort, and cost required to achieve a given level of testing quality.

Through a qualitative analysis of tool documentation and practitioner feedback, we identified the following ways in which AI-generated test artifacts contribute to increasing test efficiency.

## 1. Faster test execution with AI optimization

AI-powered tools dynamically analyze test results, historical data, and system behavior to optimize test execution sequences. Instead of running all tests indiscriminately, AI can prioritize the fastest and most relevant tests, significantly reducing overall test execution time. This is particularly beneficial in large-scale testing environments, where running a full regression suite is often impractical.

For example, some AI-powered tools use machine learning models to predict which tests are most likely to fail based on recent code changes. This allows teams to run only the most necessary tests, ensuring efficient use of computational resources and minimizing delays in continuous integration (CI) pipelines (example tools: LoadMill, ReportPortal, Testify Pulse).

Additionally, AI can parallelize test execution across multiple environments, ensuring that tests are completed faster without increasing hardware costs. Cloud-based platforms integrate AI-powered test orchestration, dynamically distributing tests to available environments based on execution priority and resource availability (example tools: LambdaTest, SauceLabs).

## 2. Reduction in manual test maintenance effort

Maintaining traditional automated test scripts is time-consuming, as even minor UI or application changes can break test cases. AI-powered self-healing test automation minimizes manual maintenance effort by automatically detecting and updating broken locators, ensuring that tests remain functional despite UI changes.

For example, self-healing AI can identify and replace obsolete element locators in test scripts without requiring human intervention. This feature significantly reduces test maintenance costs, making automation more scalable and sustainable (example tools: Testsigma, Tricentis Tosca, ACCELQ).

Furthermore, AI-powered tools can detect flaky tests—test cases that produce inconsistent results—and stabilize them by adjusting wait times, locators, or execution strategies. This increases test reliability and reduces the time spent debugging false failures (example tools: Testim, Functionize).

## 3. Minimization of test redundancy and resource waste

AI-powered test optimization eliminates redundant or duplicate tests by analyzing past execution data and removing unnecessary test cases that do not contribute to defect detection. By filtering ineffective test cases, AI ensures that only high-value tests are executed, thereby optimizing both execution time and computing resources.

For instance, AI models can cluster similar test cases, identifying and merging those that test overlapping functionalities while maintaining coverage. Additionally, AI-powered test case selection prevents re-execution of tests that have already been validated against unchanged code (example tools: Functionize, Parasoft SOAtest, TestResults.io).

Cloud-based AI solutions also monitor test execution performance over time, recommending optimizations such as removing tests that never fail or identifying areas where additional tests are needed. This intelligent test management reduces wasteful execution cycles and ensures a balanced trade-off between speed and quality (example tools: Testim, Keysight Eggplant).

## 4. Reduction of human effort in test case design

AI-generated test artifacts automate repetitive aspects of test creation, such as generating test scripts for various configurations and scenarios. This reduces the burden on test engineers, allowing them to focus on high-value exploratory testing and analysis.

For example, AI-powered test generation tools can convert natural language requirements into automated test scripts, eliminating the need for manual scripting. These tools analyze application behavior, historical defect data, and business logic to generate relevant test cases dynamically (example tools: Qodo, Testim, Virtuoso).

Additionally, AI-powered test data generation provides realistic and diverse test inputs, helping teams create better test coverage across multiple environments. By leveraging predictive analytics, AI ensures that tests target edge cases and high-risk areas more effectively (example tools: Avo Automation, Katalon).

## 5. Seamless integration with CI/CD for continuous testing

AI-powered test selection and execution integrate directly into CI/CD pipelines, ensuring that only necessary tests run for each code change, minimizing test cycle time and reducing bottlenecks in software delivery.

For instance, AI-powered impact analysis helps determine which parts of the application have changed and selects only the relevant tests to run, preventing unnecessary full-suite execution. This enables teams to execute smaller, faster, and more targeted test suites, reducing the overall time required for testing cycles (example tools: Applitools, Mabl, Tricentis Testim).

Additionally, AI-powered smart scheduling dynamically adjusts when and where tests run, optimizing resource allocation across on-premises and cloud-based environments. This allows test suites to be executed without overloading CI/CD pipelines, ensuring continuous and efficient delivery (example tools: LambdaTest, SmartBear TestComplete).

<u>A critical consideration when assessing increase in test efficiency using AI features: Effort for human oversight (peer review) of AI-generated test artifacts could negatively impact efficiency</u>

While AI-generated test artifacts greatly improve test efficiency, they always require human oversight to ensure that automated optimizations do not compromise test quality. AI-powered tools can accelerate test execution, optimize test prioritization, and reduce redundant testing, but their decision-making must be validated by test engineers to avoid missed defects, improper test exclusions, or unnecessary test simplifications.

For instance, AI-powered test selection models prioritize tests based on historical defect data and recent code changes, but they may overlook critical edge cases, low-frequency defects, or security vulnerabilities that do not frequently occur but pose significant risks. Over-reliance on AI-powered test selection without validation could lead to gaps in test coverage, missed defects, and an increased risk of software failures post-deployment.

Additionally, AI-powered test execution optimization focuses on reducing test suite runtime, but speed should not come at the cost of accuracy. If AI de-prioritizes tests due to insufficient training data, outdated models, or unexpected system behaviors, this could result in critical defects slipping through undetected. Test engineers must carefully review AI-suggested test optimizations, ensuring that high-risk areas continue to receive adequate testing and that AI does not over-prioritize execution speed at the expense of thorough validation.

Moreover, some AI-powered efficiency optimizations, such as self-healing test automation and predictive failure analysis, can sometimes apply automated fixes that are contextually incorrect. If self-healing AI incorrectly replaces UI element locators, test cases may continue to pass while actually interacting with the wrong elements, leading to false test success rates. Similarly, predictive analytics models may misinterpret system behavior anomalies, flagging harmless variations as failures or missing legitimate issues.

To mitigate these risks, organizations should adopt a hybrid AI-human validation workflow, where AI streamlines repetitive and labor-intensive tasks while test engineers validate and refine AI-powered test optimizations. By integrating human expertise with AI automation, organizations can maximize efficiency gains without sacrificing software reliability. This collaborative approach ensures that AI-powered testing remains structured, data-driven, and aligned with real-world quality assurance needs.

## 5.3 RQ3: Limitations of AI-powered testing features

While AI-powered testing tools can offer numerous benefits (as discussed in Section 5.2), they do have certain limitations that organizations must consider. Through qualitative analysis of user reviews and existing literature, we have identified several recurring themes of limitations. One would ideally expect that, as the AI technology and tools advance by time, these limitations may be partially or fully resolved by time.

1. **Lack of explainability and transparency:** Many of the AI algorithms in the tool operate as "black boxes," providing little insight into their decision-making processes, which can hinder trust and make it difficult for testers to validate AI-generated outcomes. A test engineer using the Applitools tool stated that: "*While visual AI works great, sometimes I don't understand why it marks something as a failure when it shouldn't be*". Similarly, users of Mabl have noted that its self-healing mechanism sometimes applies fixes without explaining why, making debugging challenging.

2. **Maintenance and continuous learning requirements:** AI-powered testing tools require ongoing updates and retraining to keep pace with evolving software applications. A user of the Testsigma tool remarked, "*Keeping AI models up-to-date and maintaining automated tests requires ongoing effort; the tool sometimes applies outdated learning when left unattende*d." Similarly, Tricentis Tosca has been reported to struggle with adapting to frequent UI changes unless retrained. This maintenance demand can offset the efficiency gains promised by AI automation.

3. **Complexity in implementation and integration**: Integrating AI-powered tools into existing testing workflows can be challenging. A tester using the QMetry tool has reported that: "*Adopting AI testing required process overhauls and additional team training, making it a slower transition than expected*". Similarly, an ACCELQ user noted: "*The AI's automation logic requires extensive setup before it becomes useful*". This complexity can deter teams from adopting AI-powered testing, particularly when dealing with legacy systems.

4. **Risk of overfitting and false positives:** AI models that overfit to training data may produce unreliable test results. A user of the tool DiffblueCover highlighted, "*The AI-generated test cases sometimes pass redundant assertions, flagging expected changes as issues*". Similarly, a user of the tool ReTest reported: "*The tool frequently detects UI differences that are not real defects, making it harder to focus on actual issues.*" This high rate of false positives can reduce efficiency instead of improving it.

5. **Ethical and bias concerns:** AI-powered testing tools can inherit biases from their training data, leading to skewed test prioritization. A tester using Applitools noted, "*Some visual AI tests failed disproportionately on certain interface styles, suggesting an underlying bias in its learned data.*" Similarly, a Functionize user stated, "*AI seemed to prioritize some test cases over others without clear reasoning, which raised concerns over fairness in coverage.*" Addressing these biases is critical to ensuring balanced and reliable test automation.

6. **Limited creativity and handling of edge cases**: AI-powered test case generation is less effective in identifying unpredictable user behaviors and complex business logic. A Testim user commented, "*The tool works well for common flows, but it doesn't think outside the box to explore unexpected user actions.*" Similarly, testers using Mabl mentioned that "*It automates regression tests effectively, but exploratory testing still needs human creativity.*"

7. **User control over AI model training**: Some AI-powered testing tools do not allow test engineers to train or fine-tune the AI models themselves, instead relying on vendor-controlled updates. A user of the tool Testim stated that "*We wish we had more control over training the AI for our specific use cases, rather than waiting for vendor improvements*". Similarly, in Tricentis Tosca, the AI-powered self-healing updates are applied automatically by the vendor, which limits customization for unique project needs. Organizations should carefully evaluate whether a tool provides user-accessible AI training options or if only the vendor can update the model.

8. **Feature-specific limitations**: Various data sources also discussed feature-specific limitations, which we have synthesized and discuss below:
   - Limitations of tools offering test generation using AI-powered coverage analysis: Despite its advantages, AI-powered coverage analysis can overgenerate test cases, producing redundant tests that do not necessarily improve defect detection. Some tools attempt to mitigate this by applying test prioritization techniques, but human oversight remains necessary to filter irrelevant cases and ensure that generated tests add value.
   - Limitations of tools offering test generation using AI-powered model-based testing: This approach requires accurate system models, which can be difficult for the AI tool to construct and maintain.
   - Limitations of tools offering test generation using AI-powered user behavior analytics: This approach depends on sufficient historical data, making it less effective for new applications or features without an established usage history.
   - Limitations of tools offering AI-powered visual testing: AI-powered visual testing has. One challenge is the over-detection of non-critical UI differences, where AI tools may flag minor, insignificant variations as defects. Some tools offer tunable sensitivity settings, but refining AI's ability to differentiate between acceptable and problematic changes remains an ongoing challenge. Additionally, AI-powered visual testing struggles with dynamic content, animations, and real-time UI updates, which can introduce inconsistencies in snapshot comparisons. Some tools attempt to mitigate this by employing DOM-based heuristics or context-aware comparisons, but these solutions are not yet fully reliable (tools such as: Applitools, Mabl).

As a summary of all limitations, we present in Table 2 a mapping of key limitations in AI-powered testing tools to their respective AI features. Each limitation is linked to a specific AI feature, illustrating how a certain limitation would affect different aspects of AI-powered testing. Additionally, real-world examples (derived from user reviews) point to specific AI testing tools where these issues have been observed. Understanding these limitations helps practitioners make informed decisions when adopting AI-powered software testing solutions.

**Table 4-Limitations of the AI features in the tools**

| AI feature | Key limitations | Example issues | Example tools |
|---|---|---|---|
| Generation of tests | Overgeneration of redundant tests | AI-generated test cases included excessive, repetitive scenarios | Testim |

| | Poor handling of edge cases | AI struggled to generate meaningful tests for rare defects | Diffblue Cover |
|---|---|---|---|
| Self-healing of UI tests | Incorrect locator adjustments | Self-healing modified valid locators incorrectly, breaking tests | Mabl |
| | Requires frequent retraining | Struggles with UI changes unless frequently updated | Tricentis Tosca |
| Visual testing | Over-detection of minor layout changes | AI flagged small, irrelevant UI shifts as defects | Applitools |
| | Inconsistent results across different UIs | AI's visual comparison was biased towards certain UI styles | Functionize |
| Test-result and failure analysis | Misclassification of flaky vs. real failures | AI sometimes flagged non-critical issues as severe defects | ReportPortal |
| | Lack of transparency in failure diagnostics | AI-generated root-cause analysis lacked actionable details | Testim |
| Other AI features | Risk-based test prioritization neglected uncommon but critical defects | AI-powered prioritization overly relied on past failure data, leading to under-testing of rarely failing but high-risk areas | Functionize |
| | No user control over AI model training | AI updates controlled by vendor, limiting customization | Tricentis Tosca |

## 5.4 The STR raised the need for an empirical study

The review results in Section 5 showed that, despite their promise, AI-powered testing tools exhibit varying levels of effectiveness, automation, and adaptability. The STR findings indicated that while AI-powered tools can improve test effectiveness and efficiency, their performance remains highly dependent on contextual factors, such as the complexity of UI changes, and domain-specific needs. However, since the STR was based on tool documentation, vendor claims, and practitioner reviews, we also saw the need for a direct empirical evaluation to critically observe how AI-powered features perform in real-world software testing scenarios.

To address this, we selected two representative AI-powered testing tools for an empirical study:

- Parasoft Selenic – selected for its self-healing automation feature, a widely adopted AI capability aimed at reducing test maintenance effort in Selenium-based UI testing.
- SmartBear VisualTest – chosen for its AI-powered visual testing, a critical AI-powered feature for detecting UI regressions across different environments.

These tools were selected based on three key criteria:

- Diversity of AI capabilities: They represent two distinct AI-powered testing functionalities—self-healing and visual testing—ensuring a broad assessment of AI's role in test automation.
- Maturity and industry adoption: Both tools are commercially available, actively maintained, and widely adopted by software testing teams.
- Empirical evaluability: These tools provide clear AI-powered automation that can be tested, measured, and compared against traditional test automation approaches.

The following section (Section 6) presents the empirical study methodology, detailing how the two selected test tools (Parasoft Selenic and SmartBear VisualTest) were applied to two real-world open-source systems, assessing their effectiveness, efficiency, and limitations through a structured evaluation process.

## 6 EMPIRICAL ASSESSMENT OF TWO SELECTED AI-POWERED TESTING TOOLS

This section covers the next phase of our work.

## 6.1 Design of the empirical study

### 6.1.1 Research Questions (RQs)

Let us recall from Section 3 that this entire study is guided by the three RQs. Two of the RQs (the following) were planned to be investigated in the empirical study:

- RQ2: How can AI-powered features of the tools improve effectiveness and efficiency of software testing?
- RQ3: What are the limitations of existing AI-powered testing tools, which could serve as insights for future tool enhancements and further research?

### 6.1.2 Tools under assessment

From the pool of 56 AI-powered testing tools, reviewed in Section 5, we selected two of the most mature tools, for the empirical study in this section:

- Parasoft Selenic
- Smartbear VisualTest

The primary AI feature offered by Parasoft Selenic is self-healing of UI object locators, which as discussed in Section 5.1, is automatic revision of a given Selenium test automation code when the locator encoded in the test code cannot find object in a new version of the SUT. On the other hand, the primary AI feature offered by the SmartBear VisualTest tool is visual testing, which as discussed in Section 5.1, refers to detection of visual inconsistencies or regressions in a software application's UI when the application is viewed on different settings, e.g., different screen resolutions (e.g., mobile, tablet, or desktop).

For better understanding of the empirical study and its results, we explain next some technical details of each of these two features.

### Tool 1–Parasoft Selenic: AI feature for self-healing of UI test-code

This feature is used to automatically heal test-code when changes have been made in the front-end (UI) of the SUT. The tool focuses on healing (fixing, repairing, replacing) failed "locators", developed in the context of the Selenium web-app automation framework. In Selenium test automation, locators [17] are used to identify web elements on the web page under test. The tool has a built-in AI feature, which is used to heal (repair) the locators in Selenium test-code by altering the locators which fail at runtime.

Parasoft Selenic enhances Selenium test stability through its AI-powered self-healing capabilities. When a test encounters a failure due to a broken locator or wait condition, Selenic automatically repairs the issue during execution, allowing the test to proceed without interruption. After execution, Selenic generates an HTML report detailing the encountered issues and the applied fixes.

The self-healing process involves Selenic analyzing historical test execution data to identify alternative locators for inaccessible elements. It assigns attributes such as confidence factors, weights, and stability metrics to each potential locator. These metrics are derived from previous test runs and are used to calculate the overall confidence in a locator's reliability. For instance, locators based on positional attributes may exhibit high stability but receive lower confidence scores due to their fragility in dynamic UIs. The locator with the highest confidence factor is selected to heal the test.

By integrating these AI-powered self-healing features, Parasoft Selenic could help reduce test maintenance efforts and enhance the robustness of Selenium test suites.

### Tool 2–SmartBear VisualTest: AI feature for visual testing

SmartBear VisualTest leverages advanced AI-powered computer vision techniques to detect, analyze, and highlight visual changes in web applications, ensuring UI consistency across different test runs. Unlike traditional pixel-by-pixel comparison tools, VisualTest enhances accuracy and reliability by intelligently processing both pixel data and Document Object Model (DOM) information, reducing false positives caused by minor variations such as anti-aliasing or rendering differences.

The tool functions by capturing baseline screenshots at predefined checkpoints within the functional test workflow. These baseline images serve as a reference for subsequent test executions, where VisualTest validates the visual consistency of the SUT by comparing newly captured screenshots to the stored baseline. AI-powered analysis enables the tool to intelligently detect UI alterations, distinguishing between intentional changes (e.g., design updates) and unintended regressions (e.g., misalignments, missing elements, or rendering issues).

To ensure robust comparison, VisualTest applies AI-powered heuristics to evaluate the impact of detected changes, considering both pixel variations and structural modifications within the DOM. The tool classifies differences based on severity and relevance, allowing teams to prioritize necessary fixes while minimizing test maintenance overhead.

When analyzing UI differences, SmartBear VisualTest employs a structured methodology to ensure precise comparisons across different environments. A baseline for testing is established by capturing initial screenshots, and subsequent test runs are evaluated against this reference. The AI-based comparison considers the following factors:

- Image Type: The tool supports comparison at multiple levels, including full-page screenshots, viewport-based captures, or specific UI elements.
- Operating System: The visual consistency is validated across different OS environments to detect rendering inconsistencies.

- Browser Type: Variations due to browser-specific rendering behaviors are accounted for, ensuring cross-browser UI stability.
- Screen Resolution: The tool detects UI discrepancies that arise due to differences in display resolutions, ensuring responsive design integrity.

By integrating AI-powered visual validation, SmartBear VisualTest enhances test reliability and reduces manual effort in identifying UI-related defects. It allows test engineers to automate visual testing with minimal maintenance, making it a useful tool for ensuring consistent user experiences across diverse platforms.

### 6.1.3 Systems under test (SUTs)

As the SUTs in the empirical study, we selected the following two open-source projects:

1. Spring-PetClinic [18]: A web application for managing pet clinics, developed in the Spring boot framework. Size ~ 23 KLOC
2. Java-Blog-Aggregator [19]: A software for creating and managing blogs, developed in the Spring boot framework. Size ~ 15 KLOC

Both SUTs have non-trivial size, non-trivial complexity, and have sophisticated front-end design. Since our focus in this empirical study is on UI automated testing of these two SUTs, for the sake of familiarity and context, we provide a few example screenshots from the UI of each of these two SUTs in Figure 4.

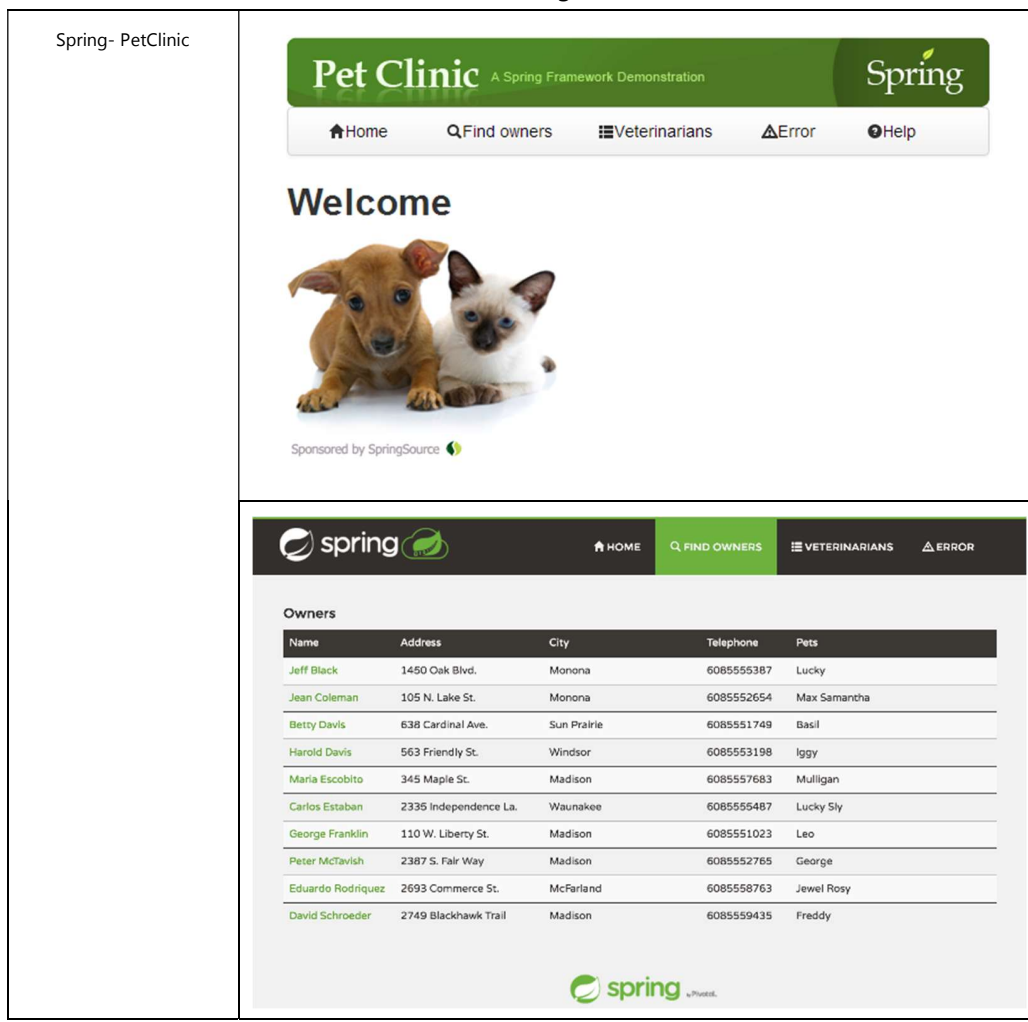

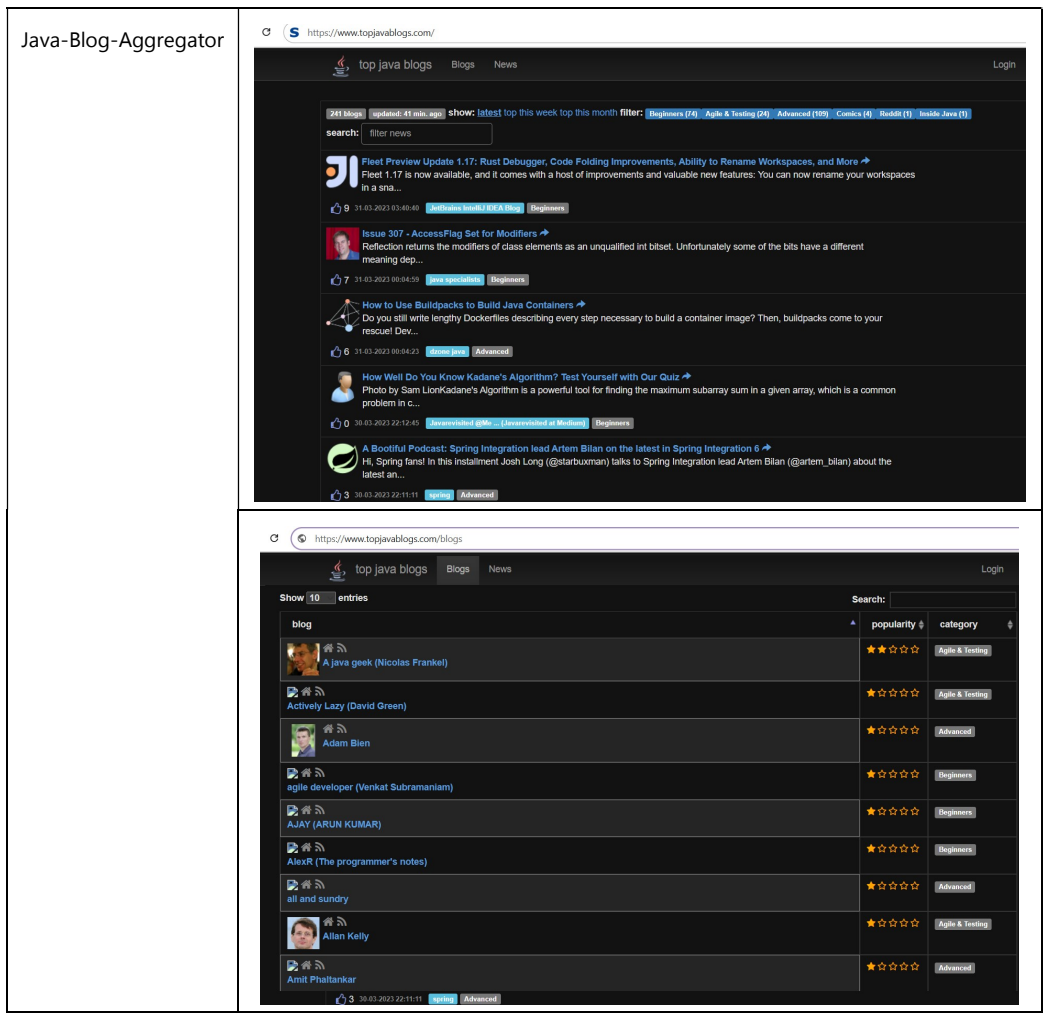

**Figure 4: Example screenshots from the UI of each of the two SUTs**

### 6.1.4 Development of a Selenium automated test-suite code-base for each of the two SUTs

To apply the two tools for testing each of the two SUTs, since no existing Selenium test suites for either of the SUT were available in their code repository, we needed to design and develop a non-trivial Selenium test suite for each of the two SUTs. We designed and developed Selenium Java test code covering all main user flows ("happy paths") in each of the two SUTs, and we provide some technical details below.

For the first SUT (Spring-PetClinic), we developed 15 Selenium test cases (methods), with test methods such as: *testAddOwner, testFindOwnerUsingSurname, testEditOwnerFirstName, testEditOwnerAddress, testAddPet, testEditPetName, testEditPetType, testAddPetVisit,* and *testExploreVeterinarians.* The code-size of the developed Selenium test-suite for the first SUT (Spring-PetClinic) was 992 LOC. We have published the full test-code-base of the two test-suites on GitHub[1]. We provide a test-code example of the PetClinic SUT in Figure 5.

For the second SUT (Java-Blog-Aggregator), we developed 18 Selenium test cases (methods), with test methods such as: *testCreateNewBlog, testEditBlogName, testEditBlogShortName, testRemoveBlog, testLogout, testAdminRemoveUser,* and *testCreateCategoryAndAssignToBlog.* The code-size of the developed Selenium test-suite for the second SUT (Java-Blog-Aggregator) was 1,804 LOC. We provide a test-code example of this SUT in Figure 6.

When developing the test code, we utilized the widely used Page-Object-Model (POM) test pattern [20]. In this pattern, each UI element (object) in a web page under test or is represented as a separate object in the test code (e.g., in the Selenium framework), encapsulating its elements and actions. The pattern offers various benefits, such as: improved test-case development and maintainability, reduced duplication in test-code, enhanced readability, and better scalability. This in turn

---

will make it easier to update test-code when the SUT's UI changes and allow for more modular and reusable test scripts across different scenarios [20].

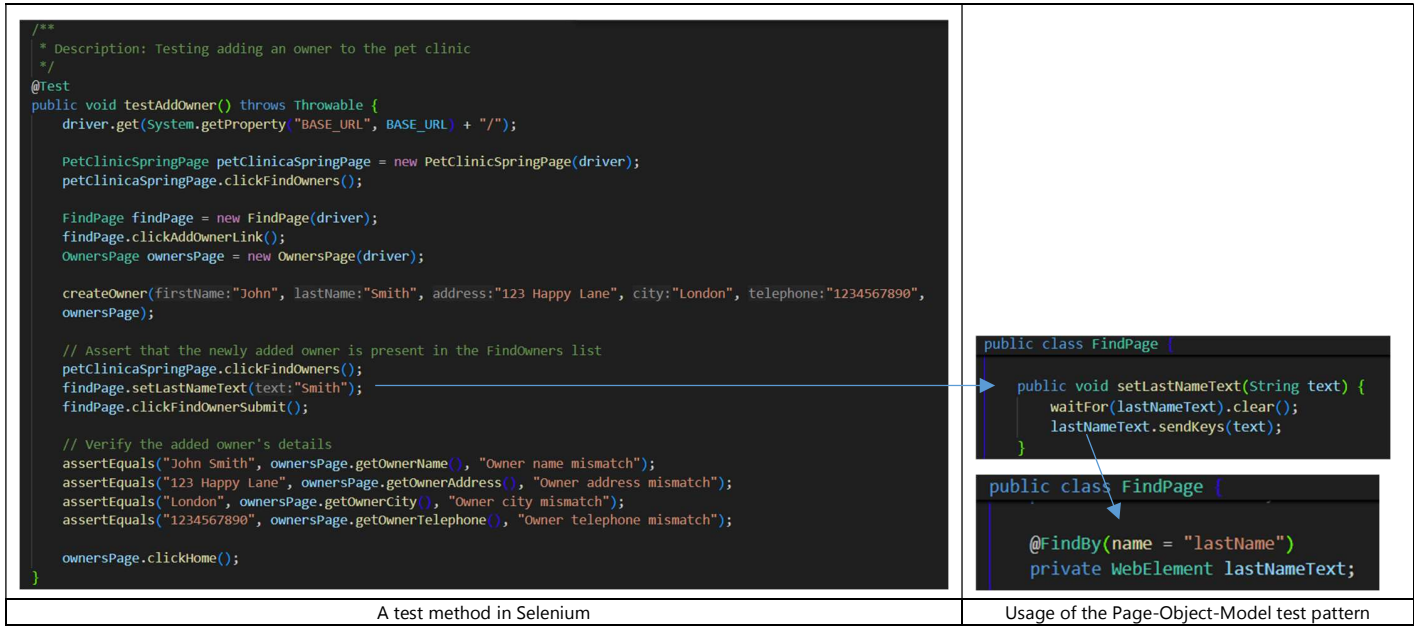

Figure 5- An example test method (left) from the Selenium automated test-suite for the SUT: Spring- PetClinic, and its Usage of the Page-Object-Model test pattern (right)

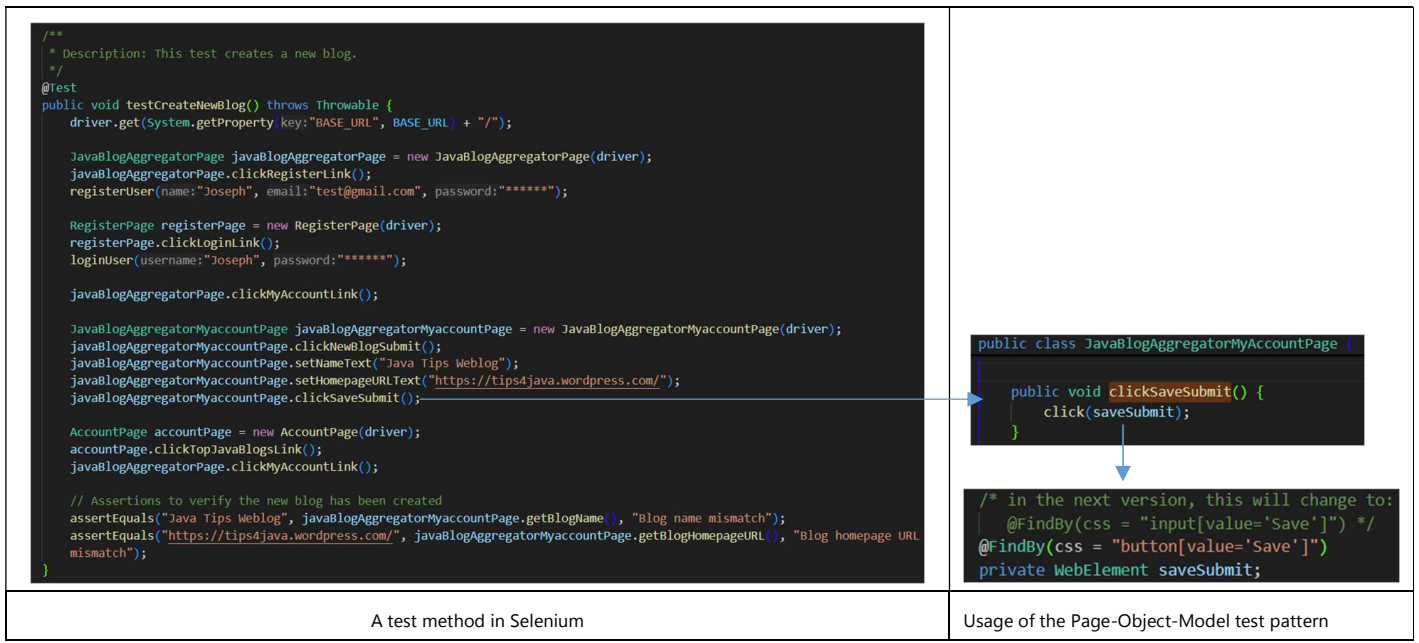

Figure 6- An example test method (left) from the Selenium automated test-suite for the SUT: Java-Blog-Aggregator, and its Usage of the Page-Object-Model test pattern (right)

### 6.1.5 Approach for assessing the effectiveness of the AI features: Mutation testing

For assessing the AI features of the two selected testing tools, our selected approach was mutation testing (intentional fault injection), which is a widely used approach for evaluation of test effectiveness [21]. We discuss below the mutation testing approach that we used for assessing each test tool (Selenic and VisualTest).

<u>Mutation-testing approach for assessing the AI self-healing feature of Parasoft Selenic:</u>

Since the AI-powered test self-healing feature focuses on the UI-object locators [17] of test-code, our mutation testing approach for this phase was to select several UI objects on each SUT, then to manually alter their attributes as mutation, and then to rerun automated test suites using the Parasoft Selenic tool, to assess whether the tool was able to identify the mutant (change) and to repair the broken tests, by suggesting to test engineer a suitable replacement for the object locator on the UI.

We did not use any automated mutation testing tool and instead conducted the mutation analysis manually, due to three reasons: (1) we did not find any suitable readily-available mutating testing tool for our tool-set setup; (2) the tool-set was already a complex set-up, and adding a new tool (mutation tool) even if we had found one, would have made the practical tool-set management and execution highly complex; and (2) we were planning to inject a dozen mutants and thus we could manage to do the mutation testing manually, in which we could have full control of the entire mutating process and analysis.

As examples of the mutation operators, for the SUT Spring-PetClinic, we manually mutated in its front-end HTML code:

```
<a class="btn btn-primary" th:href="@{/owners/new}">Add Owner</a>
```

to

```
<a class="btn btn-secondary" th:href="@{/owners/new}">Add Owner</a>
```

The above example mutation can be named as CSS-locator change, since the CSS "class" for the given hyperlink anchor object has been modified. To ensure comprehensiveness of our assessment, we made sure that all types of locator strategies in Selenium were used in the mutation analysis phase of our empirical assessment. We provide in Appendix A all the details including the list of all mutations made on each of the two SUTs, and for assessing each of the test tools. For assessing the AI self-healing feature of Parasoft Selenic, we created in total five mutants for the first SUT: Spring-PetClinic, and three mutants for the second SUT: Java-Blog-Aggregator.

### Mutation-testing approach for assessing the AI-powered visual testing feature of SmartBear VisualTest

Mutation testing was also used to assess the AI-testing feature of this tool. The plan was to make "visual" changes in the UI (front-end) of the two SUTs to see if the AI feature was able to detect the mutants.

We planned to make a variety of different UI mutations including: CSS styling changes, addition of new web elements, and moving location of page contents. For example, for the SUT Spring-PetClinic, we manually mutated in its front-end HTML code: `<div class="form-group">` to `<div class="form-group text-center">`. The above example mutation operator can be named Making-Object-Centered. Again, all the details of mutations are provided in Appendix A. For assessing the AI-powered visual testing feature of SmartBear VisualTest, we created in total five mutants for the first SUT: Spring-PetClinic, and four mutants for the second SUT: Java-Blog-Aggregator.

## 6.2 Findings of the empirical study

We present the findings of each RQs of the empirical study, in the following.

### 6.2.1 RQ1: Assessing the benefits of the tool 1 (Parasoft Selenic) – AI feature: Self-healing of UI tests

RQ2 was about the extent to which a given AI-powered feature could improve testing effectiveness and efficiency.

### Effectiveness: Does the AI feature work as expected?

In the empirical analysis, out of the five mutants in Spring-PetClinic, the tool was able to fully heal all locator failures and suggest proper replacement locators. For the 2nd SUT (java-blog-aggregator), however, the tool was only able to heal the locator failures for one of the three mutants.

We were interested to know the details of how the self-healing feature of Parasoft Selenic works (its "under the hood"). To determine replacement locators, the tool uses a combined metric called *Confidence Factor*, which is calculated from two other metrics: *Weight* and *Stability* metrics. Confidence Factor measures how accurately Selenic can repair a broken locator. The Weight represents the importance of an attribute in identifying a UI element. The Stability metric measures how consistent a locator has been in previous test runs.

Once the above metrics are calculated, the tool presents, from among all options, the locator suggestion with the highest confidence factor, since it could be the most suitable replacement for the broken locator. As an example, let us consider mutation P2 in Table 11. For that particular mutant, the calculated metrics by the tool are shown in Table 5 (these data are provided by the tool itself). From the list of available locator options, we can see that the XPath locator is the most suitable to be used to heal the broken test, due to it having the highest confidence factor which resulted from a weight of 0.98 and stability of 1.

Table 5- Confidence factor, weight, and stability of recommended locators

| Locator | Confidence factor | Weight | Stability |
|---|---|---|---|
| @FindBy(Xpath = …) | 98% | 0.98 | 1.0 |
| @FindBy(Xpath = …) | 93% | 0.98 | 0.94 |
| @FindBy(css = …) | 50% | 0.5 | 1.0 |
| @FindBy(css= …) | 50% | 0.5 | 1.0 |

In general, in this test tool, this reasoning for choosing the healed locator seems logical and systematic, and since it is presented to the test engineer during the test-repair scenario, it gives the tester confidence to use the recommended locator to heal the broken test. This satisfies the AI-explainability aspect [22] for this tool. The feature also prevents the testers from selecting an unsuitable locator that can lead again to broken (flaky) tests again, thus making the feature effective.

### Efficiency: Does the AI feature make software test engineer's job more efficient / productive?

The self-healing feature in Parasoft Selenic improves test maintenance efficiency, especially for teams using the Page-Object-Model (POM) in Selenium. In traditional workflows, identifying and updating broken locators across the POM can be tedious and time-consuming. Selenic automates this by detecting locator failures during execution and presenting suggested replacements directly in the IDE.

Instead of manually tracing locator issues, test engineers can use Selenic's unified IDE dashboard to review and apply locator fixes quickly. This reduces manual effort, lowers the risk of human error, and accelerates maintenance. The result is a more streamlined, guided process that improves engineer productivity and shortens test update cycles.

### 6.2.2 RQ1: Assessing the benefits of the tool 2 (SmartBear VisualTest) – AI feature: Visual testing

### Effectiveness: Does the AI feature work as expected?

The empirical evaluation of SmartBear VisualTest confirms that its AI-powered visual testing feature was effective in detecting UI-level issues. As shown in Table 12, the tool successfully identified all 9 injected mutants across different types of UI modifications, including layout shifts, missing components, and styling inconsistencies. This indicates that the underlying AI was capable of distinguishing visual regressions that deviate from the baseline screenshots.

Given that VisualTest performs image-based comparison with contextual awareness (e.g., resolution and browser type), the results suggest a high level of accuracy and reliability in detecting meaningful UI changes during regression testing.

### Efficiency: Does the AI feature make software test engineer's job more efficient / productive?

From the empirical assessment Smartbear VisualTest increases the efficiency of test automation, as it clearly points out to the developers which test is affected with the UI change and a link is presented in the console, which directs the user to the SmartBear VisualTest dashboard, where tester can review the detected changes. An example of this feature is shown in Figure 7. If the visual inconsistency is a regression (a defect), tester can clearly see where the issue is. Otherwise, if the visual inconsistency is intentional (and thus not a defect), the tester can "accept" the changes, and then the tool updates its baseline image.

The above AI-tester collaboration substantially decreases the time and effort the tester needs to spend finding visual inconsistency in the UI and ensures that no unintended changes would be made to the SUT.

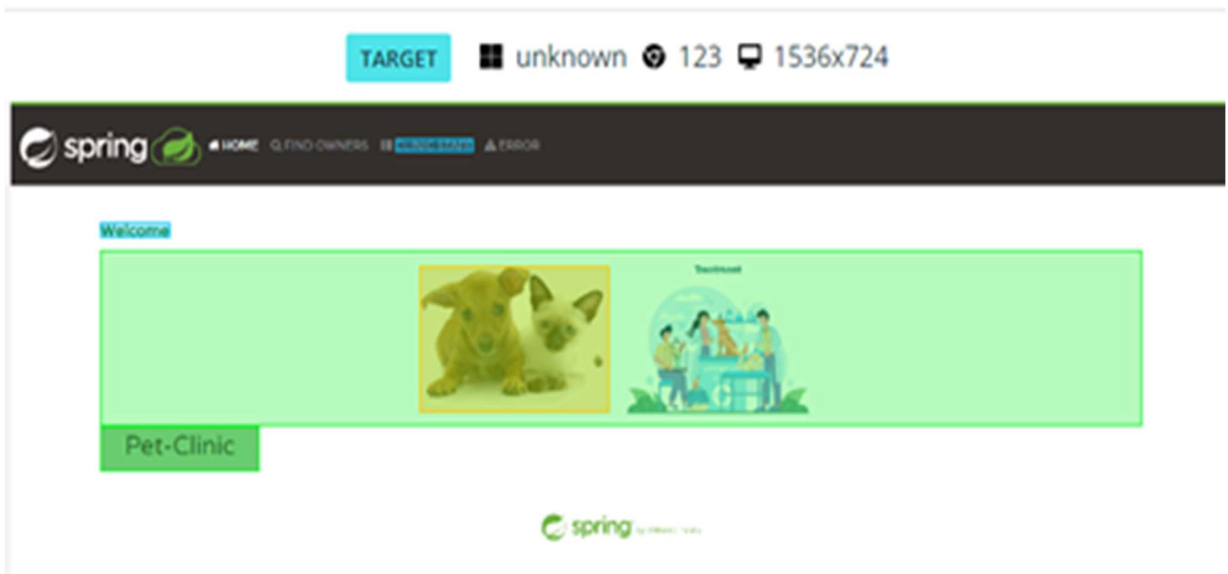

Figure 7: SmartBear VisualTest highlighting changes detected.

### 6.2.3 RQ2: Limitations of the AI features in the two tools

<u>Limitations of the AI self-healing features of tool 1 (Parasoft Selenic)</u>

When using the self-healing feature on Spring-PetClinic project, all the affected locators were changed, but when using this feature on java-blog-aggregator, which has a larger codebase, there were tests which were not able to be healed by the tool.

From Table 11, we can see that some of the locator changes were not recognized by Parasoft Selenic to provide a recommended fix. One issue is due to the tool not able to find a replacement when an element was changed from input to a button, and the locator used to identify this element was the tag name locator. As a result, all tests which used this element failed as the self-healing feature was unable to provide a suitable replacement.

Another issue which was noted was some of the tests were not healed properly even though the same locator which was used in other tests was healed. We believe this is possibly due to the self-healing feature being unable to analyze through all the affected tests. We can thus empirically conclude this tool is not fully capable of repairing (healing) all locators in all impacted tests and it is not yet fully capable of handling large test suites.

<u>Limitations of the AI visual-testing features of tool 2 (SmartBear VisualTest)</u>

In the context of this tool, when visual changes were made on the web page under test one by one, the tool was able to detect that there have been changes made. However, a limitation that was found during empirical assessment was that if multiple changes were made inside a HTML `<DIV>` element, the whole element gets highlighted and not the changes within the element.

Mutant J4 in the list of mutants in Table 12 shows that a small change that was made. However, according to the visual screenshot from the tool (Figure 8), we can see that entire HTML `<DIV>` is highlighted by the changes. This is not an issue for major UI changes, such as an addition of an image or layout change. But minor visual inconsistencies can easily be missed.

In the snapshot, there are changes made to the text of the footer which are not highlighted. This means that testers should be aware that if a visual change is small, such as addition of a character, the tester and the tool can easily miss it. In these situations, it can defeat the point of the tool. We believe that this is a minor limitation and can be overcome by having a second set of highlighters which can highlight the changes made in the DIV object itself.

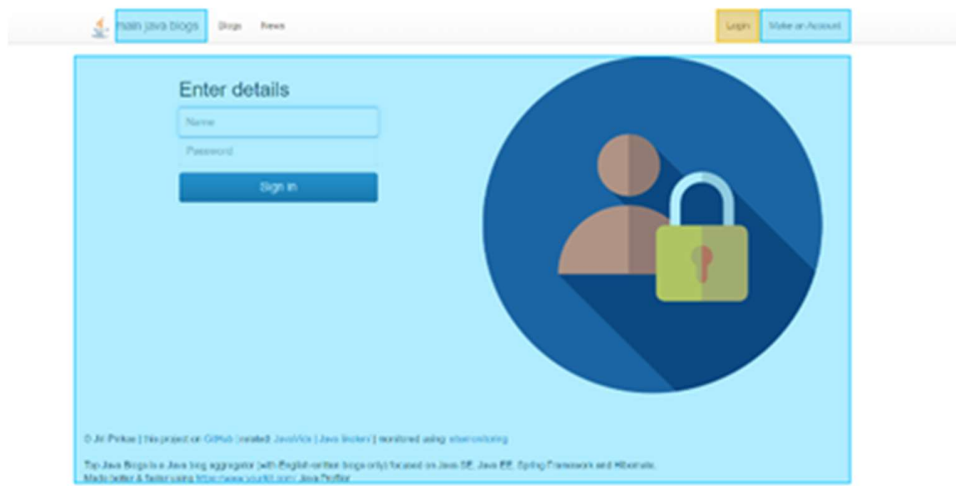

Figure 8: Highlighting lack of detail of Smartbear VisualTest when multiple changes are made.

## 6.3 Summary of the empirical study

We summarize the main findings of the empirical analysis in Table 6 and Table 7. Table 6 shows the summary of empirical analysis for the effectiveness and efficiency of using the two tools.

Table 6-Effectiveness and efficiency of the two AI-powered testing tools, as observed in the empirical study

| Tool | AI feature | (Perceived) Effectiveness of the feature | (Perceived) Efficiency of the feature |
|---|---|---|---|
| **Parasoft Selenic** | Self-healing of broken UI tests | Provides list of accurate recommendations to heal broken locators.<br>The suitable locator used to heal the test is selected based on its confidence factor, calculating from locator weight and stability metrics. | Easily replace broken locators by choosing the desired locator from recommendations and apply it to test suite via IDE.<br>Parasoft Selenic IDE interface, which clearly shows which locators failed, affected tests and list of recommended locators to fix tests. The tool navigates through all pages in the test suite and fixes all broken locators with a locator user selected from the list of recommendations. |
| **SmartBear VisualTest** | Visual testing | Able to detect UI changes such as moving content<br>AI-based computer-vision is used to compare the screenshot with the baseline image. Tool freezes the moving contents at same point in its cycle from capture to capture to ensure actual changes in SUT are detected. | Test fails if visual inconsistencies are detected. The tool shows the issue to the user.<br>Dashboard clearly shows the visual inconsistencies using colored boxes, different ways to view change such as, side by side comparison, drag and compare. Changes can be accepted (updates baseline image) or rejected by the user. |

Table 7 presents a summary of the limitations of each feature that were discovered in the empirical study. To summarize, it is valid to state that the AI-powered testing tools, Parasoft Selenic and SmartBear VisualTest do increase the efficiency and effectiveness of test automation. Self-healing feature of Parasoft Selenic reduces the time for test maintenance, while SmartBear Visual Test enables detection of visual defects more efficiently.

Table 7-Limitations of the two selected AI-powered testing tools, as observed in the case study

| Tool | AI feature | (Perceived) Limitations of feature | Possible reason(s) leading to the limitation | Suggested improvements for the AI feature |
|---|---|---|---|---|
| **Parasoft Selenic** | Self-healing of broken UI tests | Unable to heal certain locators, and could not heal locators in certain pages of the SUT | It is possible that the AI model behind this feature is not advanced enough to always find an alternative to a failed locator, and not yet capable of examining through all pages in test-suites developed using the page-object pattern. | We could expect that the limitation will improve in the near future, by the tool vendor, due to the AI model getting further trained on more extensive datasets |
| **SmartBear VisualTest** | Visual testing | Unable to mark visual inconsistencies individually. Only the area of change is highlighted not the | . If one change is made at a time, the tool can recognize the change, but when multiple changes are made, it cannot locate each individual change. This shows that the functionality is not yet present in the tool, | We could suggest that the functionality of highlighting each individual change made on UI is added to the tool. Without such a detailed feature, certain |

| | | individual changes within the highlighted area | and most possibly it is not an issue with AI model. | visual inconsistencies can be missed by the tester. |
|---|---|---|---|---|

## 6.4 Generalizability of the findings from our empirical study

It is important to discuss the generalizability of our findings beyond the two selected test tools, and the two open-source SUTs used in this study. The empirical study assessed the effectiveness and efficiency of two AI-powered testing tools—Parasoft Selenic (self-healing test automation) and SmartBear VisualTest (AI-powered visual testing)—when applied to two open-source systems. The results demonstrate that AI-powered automation can significantly reduce test maintenance effort and improve defect detection, validating findings from the Systematic Tool Review (STR). However, the study also revealed certain limitations, such as partial failure handling in locator self-healing and inconsistent granularity in visual defect identification.

While this study was conducted on two open-source software systems (Spring-PetClinic and Java-Blog-Aggregator), the findings can be generalized to enterprise testing environments in several ways. First, the selected open-source projects mimic real-world software architectures, incorporating complex UI structures, database interactions, and multi-layered business logic—features also present in enterprise applications. Second, the AI-powered features evaluated in this study (self-healing and visual testing) address challenges common across software domains, including frequent UI changes, test maintenance overhead, and the need for automated regression testing. Enterprise systems with large-scale UI components and frequent release cycles are particularly well-suited for AI-powered test automation.

However, enterprise applications often involve greater system complexity, including legacy components, highly customized workflows, security constraints, and large-scale CI/CD pipelines. While the observed benefits of AI-powered testing tools—such as reducing test maintenance effort and improving test execution efficiency—are likely transferable to enterprise contexts, further empirical evaluations on larger-scale applications would help validate these findings. Future studies could explore how AI-powered testing tools perform in large-scale enterprise CI/CD pipelines, assess their scalability in high-traffic environments, and investigate their adaptability to domain-specific testing needs (e.g., healthcare, finance, and embedded systems testing).

The findings from our empirical study reinforce key insights from the Systematic Tool Review (STR), demonstrating that AI-powered testing tools provide notable efficiency gains but still require human oversight to handle complex test scenarios and reduce false positives. These results highlight critical areas for future AI-powered test automation improvements, including better explainability, improved failure handling, and enhanced adaptability to diverse testing environments. The following section discusses the broader implications of these findings, potential limitations of our study, and directions for future research.

## 7 DISCUSSIONS, CONCLUSIONS AND FUTURE WORK DIRECTIONS

## 7.1 Fully autonomous software testing remains a distant goal

The concept of fully autonomous software testing has been a long-standing vision in the testing community, where AI and automation technologies could manage the entire testing lifecycle—from test generation to execution and defect analysis—without "any" human intervention. However, despite recent advances in AI-powered testing, full autonomy remains a distant goal.

Let us recall the six levels of autonomy in software testing, which was shown in Figure 1 in Section 2.1. While Levels 1-3 are already present in modern AI-powered testing tools (e.g., self-healing automation, AI-powered test generation), Levels 4-5 remain largely theoretical due to fundamental limitations in AI's ability to understand business logic, edge cases, and contextual nuances.

Several limitations (challenges) currently hinder the progression toward Level-5 autonomous testing:

- Limited AI Reasoning and Context Awareness: AI models lack deep domain understanding and struggle with complex business logic, security policies, and compliance requirements.
- False Positives and False Negatives: AI-powered defect detection tools frequently misclassify minor UI shifts as bugs while missing critical functional regressions.

- Lack of Explainability in AI Models: Many AI-powered testing tools operate as black boxes, making it difficult for testers to trust AI-generated test cases and defect classifications.
- Evolving Software Environments: AI models require continuous retraining to keep up with UI updates, backend changes, and evolving system behaviors, increasing maintenance overhead.
- Human Judgment and Exploratory Testing: AI lacks the ability to think critically, adapt test cases to new requirements, or explore an application in unpredictable ways like human testers.

The conclusion is that AI is an assistant, not a replacement for human test engineers. The findings from our Systematic Tool Review (STR) and empirical study confirm that full AI autonomy in software testing remains a distant goal. While AI-powered tools provide significant efficiency gains (e.g., self-healing automation, AI-powered visual testing), they still require human testers to review, refine, and validate AI-generated test artifacts.

## 7.2 Implications of findings for practitioners (test engineers and QA teams)

AI-powered testing tools provide significant benefits in reducing test maintenance, improving defect detection, and accelerating test execution. However, practitioners should be aware of the limitations of these tools and adjust their testing strategies accordingly.

AI-powered testing tools are best used as intelligent assistants rather than full replacements for human testers. While AI can optimize workflows and enhance test efficiency, human oversight remains essential to refine AI-generated artifacts, interpret results, and mitigate risks.

In Table 8, we provide key takeaways for test engineers and QA teams.

Table 8- Key takeaways for test engineers and QA teams

| Area | Implication |
| --- | --- |
| AI-Powered Test Generation | AI-generated tests can expand coverage and speed up test creation, but engineers must review AI-generated test cases to ensure they align with business requirements. |
| Self-Healing Test Automation | Self-healing features reduce maintenance effort but may not handle complex UI changes. Testers should validate AI-suggested locator updates before applying them. |
| AI-Powered Visual Testing | While AI improves UI validation, it may generate false positives for minor UI shifts. Testers must configure sensitivity settings and manually review flagged defects. |
| Test Result & Failure Analysis | AI-powered failure detection can prioritize debugging efforts, but testers should still investigate false positives and misclassifications manually. |
| AI in CI/CD Pipelines | AI-powered test selection reduces execution time but should not be blindly trusted—teams must continuously monitor AI's decision-making and fine-tune test prioritization rules. |

## 7.3 Implications of findings for tool developers

The study identified several challenges and limitations in AI-powered testing tools, presenting opportunities for tools improvement. Tool vendors and AI researchers could focus on enhancing AI models and reducing tool adoption barriers. In Table 9, we provide the key areas for tools improvement.

Table 9- Key areas for tools improvement

| Challenge | Improvement needed |
| --- | --- |
| False positives in AI-powered visual testing | Improve context-aware AI models that can distinguish meaningful UI defects from minor rendering changes. |
| Lack of explainability in AI-powered test selection | Provide explainable AI (XAI) features that justify why certain tests were selected, prioritized, or de-prioritized. |
| Limited adaptably of self-healing locators | Enhance AI to better generalize across different UI frameworks and improve handling of dynamic applications. |
| AI model retraining requirements | Develop adaptive AI models that can continuously learn from new test execution data without frequent manual retraining. |
| Scalability in large-scale CI/CD pipelines | Optimize AI-powered testing tools for high-volume test execution environments, reducing resource consumption while maintaining accuracy. |

For AI-powered testing tools to gain wider industry adoption, vendors must focus on improving AI explainability, reliability, and adaptability. Enhancing human-AI collaboration—rather than aiming for full automation—will make AI-powered tools more effective and trustworthy in real-world testing environments.

## 7.4 Potential threats to validity

We discuss in this section the potential threats to validity of the study and the actions we took to minimize or mitigate them, based on the guidelines in software engineering [23].

Let us recall from the earlier section that this study had two components: (1) A systematic review of AI-powered testing tools, categorizing their key AI features; (2) An empirical study of two selected tools. We present in Table 10 the potential threats to validity of each study component, and the mitigation strategies that we have used. We categorize threats into four validity aspects: internal validity, external validity, construct validity, and conclusion validity.

**Table 10- Potential threats to validity of the study, and the mitigation strategies**

| Study component | Threat type | Threat description | Mitigation strategy |
|---|---|---|---|
| Systematic tool review | Internal validity | Selection bias in tool inclusion – Some AI-powered tools may not explicitly label their AI capabilities, leading to potential omissions. | Conducted multiple search iterations using different AI-related keywords (e.g., "AI-powered testing," "machine learning in testing"). Reviewed vendor documentation, product sheets, and independent software directories. |
| | | Accuracy of extracted data – Tool documentation, vendor reports, and practitioner reviews may provide incomplete or misleading information. | Cross-verified data from multiple independent sources (official tool documentation, G2.com reviews, and industry reports) to ensure reliability. |
| | External validity | Generalizability of findings – STR findings may not remain relevant as AI-powered testing evolves. | Explicitly documented inclusion criteria and categorized tools based on AI functionalities, allowing future studies to update findings. |
| | | Applicability to enterprise environments – STR focused on publicly available tools, raising concerns about relevance to large-scale enterprise applications. | Selected tools with active industry adoption to ensure findings remain relevant to both small-scale and enterprise testing environments. |
| | Construct Validity | Distinguishing AI-powered features from rule-based automation – Some tools may claim AI capabilities without true AI implementation. | Carefully analyzed AI implementation details, ensuring tools explicitly leveraged ML, NLP, or computer vision, rather than simple rule-based automation. |
| | | Varying degrees of AI automation – Some tools require significant human intervention despite claiming AI-powered functionality. | Categorized tools based on their level of automation, distinguishing between fully automated AI features and AI-assisted functionalities requiring tester involvement. |
| | Conclusion Validity | Reliance on vendor claims and user reviews – Without direct experimental validation, there is a risk of bias in reported tool benefits. | Triangulated data from multiple independent sources to avoid sole reliance on vendor claims. Conducted an empirical study to validate STR findings. |
| Empirical study | Internal validity | Selection bias in tool choice – Only two tools (Parasoft Selenic and SmartBear VisualTest) were selected, limiting the scope of AI feature evaluation. | Selected tools based on maturity, feature richness, and industry adoption to ensure broader applicability. Focused on AI capabilities rather than tool-specific implementations, making insights transferable to similar AI-powered testing tools. |
| | | Mutation testing approach – Manual fault injection may not fully replicate all real-world test scenarios. | Selected diverse mutation types (locator changes, CSS modifications, structural UI updates) to ensure a broad evaluation of AI-powered functionalities. |
| | External validity | Generalizability beyond open-source systems – The empirical study used two open-source projects, which may differ from large-scale enterprise applications. | Selected realistic and structured open-source projects with complex UI components and database interactions to reflect real-world software architectures. Future research should extend evaluation to large-scale enterprise systems. |
| | Construct validity | Testing conditions vs. real-world enterprise usage – Industrial applications often involve larger test suites, complex workflows, and CI/CD integration. | Designed Selenium-based test cases to represent typical UI test scenarios (form interactions, navigation flows, input validation). Suggested future work on large-scale enterprise CI/CD integrations. |

| | Conclusion validity | Subjectivity in interpreting tool effectiveness – Some findings, such as ease of use and efficiency gains, may be qualitative. | Documented quantifiable metrics (e.g., self-healing success rates, defect detection accuracy). Included real-world user feedback from G2.com to validate observations. Compared AI-powered results to traditional test automation to ground findings in real-world test scenarios. |
|---|---|---|---|

## 7.5 Conclusions and future work directions

AI-powered software testing is rapidly gaining adoption in the industry, with projections indicating that 80% of enterprises will integrate AI-augmented testing tools by 2027. This study provides a comprehensive evaluation of AI-powered testing tools by conducting a Systematic Tool Review (STR) of 56 tools and an empirical assessment of two representative tools, Parasoft Selenic and SmartBear VisualTest.

The STR categorized AI-powered tools based on self-healing automation, AI-powered test generation, visual testing, test prioritization, and predictive analytics. Findings show that AI-powered features enhance test automation efficiency, reduce maintenance costs, and improve defect detection rates. However, challenges such as false positives, contextual limitations, and over-reliance on predefined AI models remain barriers to their full potential.

The empirical study provided an in-depth assessment of two AI-powered tools on real-world open-source systems. Results demonstrate that AI-powered self-healing automation reduces test maintenance effort, while AI-powered visual testing improves UI validation accuracy. However, limitations such as partial failure handling in locator self-healing and lack of granular highlighting in visual testing indicate areas needing improvement.

A key takeaway is that AI-powered testing tools still require human oversight. While AI can automate repetitive tasks and optimize test execution, human testers must review AI-generated test artifacts to ensure correctness, contextual relevance, and adaptability to changing software conditions.

AI-powered software testing offers significant potential to enhance test automation, defect detection, and software quality assurance. However, AI-powered tools are not a "silver bullet"—they require continuous improvements in AI models, better integration with software development lifecycles, and stronger collaboration with human testers. As AI in software testing continues to evolve, future advancements in AI-powered test optimization, risk-based prioritization, and explainable AI will play a critical role in shaping the next generation of software quality assurance practices.

This study highlights several avenues for future research and tool enhancement in AI-powered software testing:

- Enhancing AI explainability and transparency: Many AI-powered testing tools function as "black boxes," providing limited visibility into AI-powered decisions. Future research should focus on developing explainable AI (XAI) models that offer testers clear rationales for test selection, failure detection, and self-healing decisions.
- Improving AI model generalization: Current AI-powered testing tools often struggle with handling complex UI changes, business logic updates, and dynamic application states. Enhancing AI model training using more diverse datasets and real-world defect patterns could improve their adaptability to different software environments.
- Advancing AI-powered test prioritization: AI-based test selection and execution optimization have shown promise, but risk-based test prioritization algorithms need refinement. Future research should explore context-aware test prioritization models that dynamically adapt test selection based on evolving software risks and changes in code complexity.
- Reducing false positives and enhancing defect localization: One limitation of AI-powered tools is their tendency to over-detect minor UI changes or misclassify non-critical issues as defects. Research should focus on developing hybrid AI models that combine computer vision, contextual analysis, and historical defect data to minimize false positives and improve defect localization accuracy.
- Integrating AI-powered testing with CI/CD pipelines: While AI-powered tools can improve test execution speed, their integration with modern DevOps workflows remains an open challenge. Future work should focus on seamless AI-powered test orchestration, where AI dynamically schedules, executes, and refines test cases within continuous integration/continuous deployment (CI/CD) pipelines.
- Exploring AI for exploratory and security testing: Current AI-powered tools focus on functional and regression testing, but AI-powered approaches for exploratory testing, security vulnerability detection, and robustness testing remain underexplored. Future research could investigate agentic AI models capable of autonomously exploring application behaviors, identifying security flaws, and executing adaptive security tests.

- Human-AI collaboration in software testing: AI-powered tools are not yet fully autonomous, and human oversight remains critical. Future research should explore how test engineers and AI can collaborate effectively, with AI acting as an intelligent assistant rather than a replacement for human testers.

## ACKNOWLEDGEMENT

Vahid Garousi was partially funded from the HIVEMIND project funded by the European Union under Grant Agreement nr. 101189745.

# APPENDIX A-DETAILS OF THE EMPIRICAL RESULTS

### Table 11-Empirical assessment results for the tool: Parasoft Selenic

| SUT | Changes (mutations) in the front-end of the SUT | | | Mutation operator (change in type of Selenium locator) | Did any UI test fail on the mutant? | Was the AI tool able to identify the issue and fix the broken test? | What approach did the AI tool take to fix (repair) the broken test? |
|---|---|---|---|---|---|---|---|
| | ID | Before mutation | After mutation | | | | |
| Spring-PetClinic | P1* | `<a class="btn btn-primary" th:href="@{/owners/new}">Add Owner</a>` | `<a class="btn btn-secondary" th:href="@{/owners/new}">Add Owner</a>` | CSS locator | Yes | Yes | Altered the locator to link text, using "Add Owner" |
| | P2 | `<button type="submit" class="btn btn-primary">Find Owner</button>` | `<button type="submit" class="btn btn-primary">Search for Owner</button>` | XPath locator | Yes | Yes | Altered the locator to XPath |
| | P3 | `<a id="editLink" th:href="@{__${owner.id}__/edit}" class="btn btn-primary">Edit Owner</a>` | `<a id="editOwnerLink" th:href="@{__${owner.id}__/edit}" class="btn btn-primary">Edit Owner</a>` | ID locator | Yes | Yes | Altered the locator to link text, using "Edit Owner" |
| | P4 | `<a th:href="@{__}" class="btn btn-primary">Add New Pet</a>` | `<a th:href="@{__}" class="btn btn-primary">Register Pet</a>` | Link text locator | Yes | Yes | Altered the locator, to ByClass |
| | P5 | `<button th:with="text=${owner['new']} ? 'Add Owner' : 'Update Owner'" class="btn btn-primary" th:text="${text}">Add Owner</button>` | `<button th:with="text=${owner['new']} ? 'Add Owner' : 'Update Owner'" class="carousel-dark" th:text="${text}">Add Owner</button>` | Class name locator | Yes | Yes | Altered the locator to XPath |
| Java-Blog-Aggregator | J1 | `<li><a href="/logout" th:text="${'Logout'}"></a></li>` | `<li><a href="/logout" th:text="${'Sign Out'}"></a></li>` | Link text locator | Yes | No | Altered the locator to link text locator, in some tests. But the other tests were not changed and were still broken. |
| | J2 | `<input type="submit" class="btn btn-primary" value="Save" />-->` | `<button type="submit" class="btn btn-primary" value="Save">Save</button>` | Tag name and CSS locator | Yes | No | Altered the locator, to ById, but that did not work either |
| | J3 | `<a th:href="${'/blog-form?blogId=' + blog.id}" class="btn btn-primary">edit</a>` | `<a th:href="${'/blog-form?blogId=' + blog.id}" class="btn btn-primary">Edit Blog</a>` | Link text locator | Yes | Yes | Altered the locator, to partial link text |

## Table 12-Empirical assessment results for the tool: SmartBear VisualTest

| SUT | Changes (mutations) in the front-end of the SUT | | | Mutation operator (type of visual change) | Did any UI test fail on the mutant? | Was the AI tool able to identify the issue? |
|---|---|---|---|---|---|---|
| | ID | Before | After | | | |
| Spring-PetClinic | P1* | `<div class="form-group">` | `<div class="form-group text-center">` | Shifted the button to the center of the screen, by changing its CSS class type | Yes | Yes |
| | P2 | `<div class="col-md-12">`<br>`    `<br>`</div>` | `<div class="col-md-12">`<br>`    <centre></centre>`<br>`</div>` | Shifted the image location to center | Yes | Yes |
| | P3 | `<a id="editLink" th:href="@{…}" class="btn btn-primary">Edit`<br>`        Owner</a>` | `<a id="editLink" th:href="@{…}" class="btn btn-secondary">Edit`<br>`        Owner</a>` | Changed styling of the button by changing its CSS class type | Yes | Yes |
| | P4 | `th:replace="~{fragments/inputField :: input ('Last Name', 'lastName', 'text')}" />` | `th:replace="~{fragments/inputField :: input ('Surname', 'lastName', 'text')}" />` | Changed the wording of the text input box | Yes | Yes |
| | P5 | `None` | `<div id="box">Pet-Clinic</div>` | Added a text division | Yes | Yes |
| Java-Blog-Aggregator | J1 | `N/A` | `<div class="col-md-6">`<br>`  <img class="img-responsive"    src="…"`<br>`th:src="@{…}" style="max-width: 100%; height: auto;">`<br>`</div>` | Added an image to the login page | Yes | Yes |
| | J2 | `configuration.setBrandName("top java blogs");` | `configuration.setBrandName("main java blogs");` | Changed the brand name | Yes | Yes |
| | J3 | `<li th:class="${current == 'register' ? 'active' : ''}"><a href="/register">Register</a></li>` | `<li th:class="${current == 'register' ? 'active' : ''}"><a href="/register">Make an Account</a></li>` | Changed text of list item | Yes | Yes |
| | J4 | `<p> Made better & faster using <a href="https://www.yourkit.com/">https://www.yourkit.com/</a> Java Profiler </p>` | `<p> Made better & faster using <a href="https://www.yourkit.com/">https://www.yourkit.com/</a> Java Profiler. </p>` | Added full stop at end of sentence | Yes | Yes |

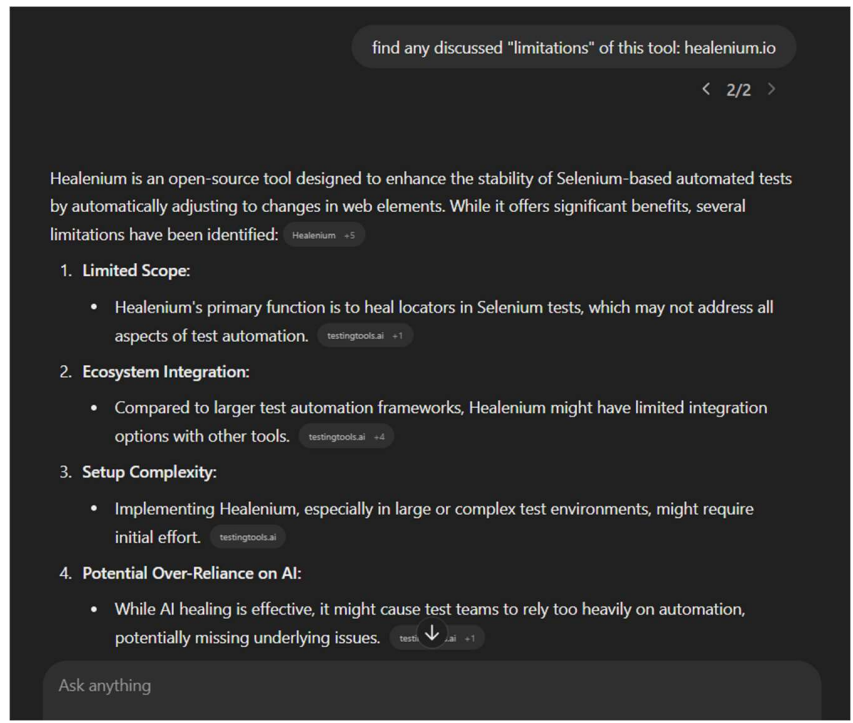

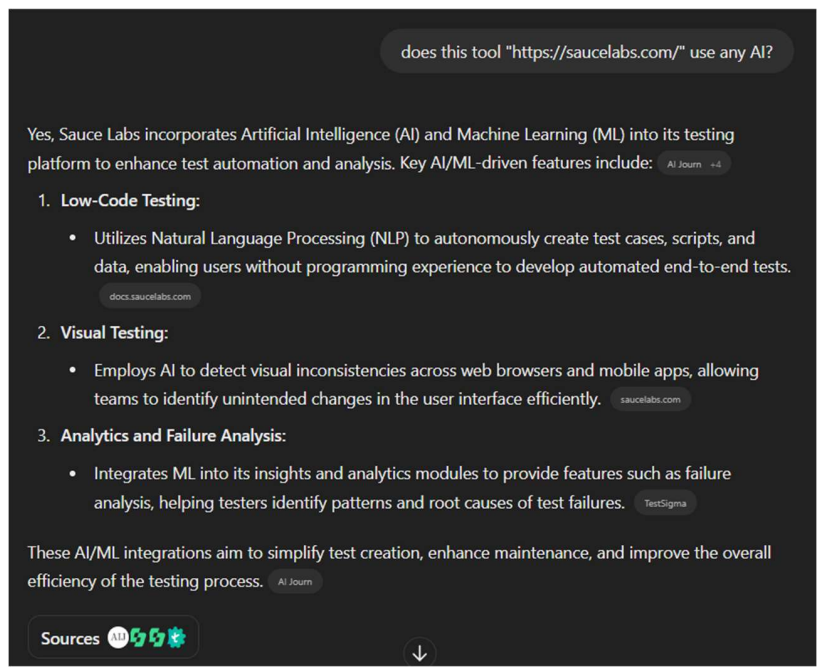

When the tools website did not clearly mention its AI features:

find the AI feature in this tool: https://www.tricentis.com/

Tricentis integrates Artificial Intelligence (AI) across its suite of software testing tools to enhance automation, efficiency, and quality. Key AI-powered features include: tricentis.com +3

1. **Tosca Copilot:**
   - A generative AI assistant that boosts productivity and application quality throughout the testing lifecycle. It enables users to quickly find, understand, and optimize test assets via a user-friendly chat interface. tricentis.com

2. **Vision AI:**
   - This technology mimics human vision to automate testing for applications that were previously challenging to automate, such as those running on remote desktops. It can even create test automation based on design mockups before any code is written, allowing for earlier testing in the development lifecycle. tricentis.com

3. **qTest Copilot:**
   - A generative AI test generation assistant that drafts test cases with steps and expected results by analyzing requirements in qTest. It accelerates test case creation, broadens testing scope, and enhances release quality. tricentis.com

4. **AI-Powered Quality Engineering Solutions:**
   - Tricentis offers AI-augmented tools designed  to accelerate application modernization, enhance

Ask anything

Table 3 –The AI testing tools and classification of their features

| Tool names | AI features of the tools | | | | | If test generation, what approach is used? | | | | Level (type) of tests generated / repaired | | | |
|---|---|---|---|---|---|---|---|---|---|---|---|---|---|
| | Test generation | Self-healing of UI tests | Visual testing | Test-result and failure analysis | Other AI features | Low (no) Code, using NLP | Low (no) Code, using Record / Playback | Code-coverage analysis | Other | Unit tests | API (integration) tests | UI (system) tests | Only test cases, no test code |
| # of tools under each category → | 43 | 32 | 16 | 11 | 23 | 20 | 9 | 13 | 3 | 16 | 3 | 20 | 2 |
| ACCELQ | x | x | | | | x | | | | | | x | |
| Amazon CodeWhisperer | x | | | | | x | | | | x | | | |
| Applitools | | x | x | | AI is used to ignore dynamic content like ads, reducing unnecessary tests | | | | | | | | |
| Appsurify - TestBrain | | x | | | Uses AI to provide alerts on risky code changes to catch defects early. | | | | | | | | |
| Appvance | x | x | | | | | | x | | | | x | |
| Aqua | x | | | | Test prioritization | | | | | | | x | |
| Autify | x | x | x | | | | | | | | | x | |
| Avo Automation | x | x | | | • Change-impact analysis<br>• Automated test-data generation | x | | | | x | | | |
| BaseRock (formerly: Sapient) | x | x | | | Creating realistic synthetic test data | | | x | | x | x | | |
| BrowserStack - Percy | | | x | | | | | | | | | | |
| Cloud QA | x | x | | | | | x | x | | | | x | |
| Code Test Generator | x | | | | | | | x | | x | | | |
| qodo.ai (formerly: Codium.io) | x | x | | | • Agentic test workflows: Chat-based, guided test generation for tailored testing<br>• AI-powered test suite extension and validation<br>• Context-Aware test-code customization | | | x | | x | x | | |
| CommandDash.io (formerly: welltested.ai) | x | | | | | | | x | | x | | | |
| Copado - Robotic Testing | x | x | | | Predictive analytics for detailed reporting | x | | | | | | x | |
| diffblue Cover | x | | | | Provides feedback on testability of code | | | x | | x | | | |
| digital.ai | x | x | | | | x | x | | | | | x | |
| Functionize | x | x | x | | A gen-AI feature named testASSIST enables non-technical users to manipulate testing scenarios and streamline their test case creation process | x | | | | x | | x | |
| Gemini Code Assist (formerly Duet AI) - Google | x | | | | | | | x | | x | | | |
| Github CoPilot | x | | | | | x | | x | | x | | x | |
| Healenium | | x | | | | | | | | | | | |
| iHarmony | x | x | | | | | x | | | | | | |
| IntelliJ IDEA | x | | | | | | | | | x | | | |
| Katalon | x | x | x | x | | x | | | | | | web apps | |
| Keysight - Eggplant Test | x | | x | | AI-powered exploratory testing | x | | | Model-based testing | | | x | |
| Kobiton | x | x | x | | | | | x | | | | mobile apps | |
| LambdaTest HyperExecute | | | x | | An AI feature named Test Orchestration: automatically groups and distributes tests intelligently across different testing environments | | | | | | | | |
| Launchable | | | | x | | | | | | | | | |
| LoadMill | x | x | | x | | | | | User behavior (operational profile, logs) | | | x | |
| Mabl | x | x | | | | x | | | | | | | |
| Machinet AI Unit Tests (a plugin for IntelliJ) | x | | | | --AI describes what each test method is doing<br>--AI generates mocks, objects, and non-trivial assert statements<br>--AI decides whether to use mocks or not. | | | x | | x | | | |
| Parasoft JTest | x | | | | | | | x | | x | | | |
| Parasoft Selenic | | x | | x | | | | | | | | | |
| Parasoft SOAtest | x | | | | | x | | | API logs (recorded traffic) | | x | | |
| pCloudy | x | x | x | | Observability agent, for full testing visibility | x | | | | | | x | |
| Perfecto | | x | | | | | | | | | | | |
| ReportPortal | | | | x | | | | | | | | | |
| retest | x | | x | | | | x | | | | | x | |
| SauceLabs | x | x | x | x | | x | | | | | | x | |
| Smartbear - VisualTest | | | x | | AI-powered object recognition | | | | | | | | |
| Smartbear - TestComplete | | x | | | | | | | | | | | |
| Sofy Co-Pilot | x | | x | x | --Hyper-intelligent (optimized) regression testing<br>--Possibility of further training of the built-in AI model | x | | | | | | mobile apps | |
| taskade | x | | | | | | | | | | | | x |
| Tabnine | x | | | | | | | | | x | | | |
| Test Grid | x | x | | x | | x | | | | x | | | |
| Testify - Pulse | x | | | x | --AI-hierarchical clustering: groups API behaviors based on similarities, identifying both documented and undocumented patterns. It ensures each pattern is tested, improving coverage and detecting edge cases.<br>--AI dependency analysis: Maps system component relationships for better test generation | | | x | | | x | | |
| TestingWhiz | x | x | x | | | | x | | | | | | |
| TestResults | | x | x | | --AI Virtual user tests the software as a user would.<br>--Visual hints: AI-powered spatial algorithm to detect interaction points without predefined hints, enabling adaptive and robust automated testing. | | | | | | | | |
| TestRigor | x | x | | | | x | | | | | | | |
| TestSigma | | x | | | | x | | | | | | | |
| Tricentis Testim | x | x | | | --Learns user flows, recognizes repeated sequences and suggests reusable elements.<br>--Captures screenshot of every test step and compares to previous screenshots to identify what has been changed, to suggest appropriate test cases.<br>--AI helps in the development of well-architected, clean tests that optimize reuse and minimize maintenance. | | | x | | | | x | |
| Tricentis Tosca | x | x | x | | --A feature named RiskAI analyzes changes in SAP systems to identify high-risk areas needing prioritized testing, reducing release scope while achieving risk coverage | x | x | | | | | | x |
| Unit-test.dev | x | | | | | | | | | x | | | |
| Virtuoso | x | x | | | Test data generation | x | | | | | | web apps | |
| Webo.ai (by Webomates) | x | x | | | | | x | | | | | web apps | |
| Workik | x | | | | | | | x | | x | | | |